\documentclass{article}

\usepackage{microtype}
\usepackage{graphicx}
\usepackage{subfigure}
\usepackage{booktabs} 
\usepackage{tabularx}
\usepackage{multirow}
\usepackage{array}
\usepackage{enumitem}

\usepackage{hyperref}

\usepackage[accepted]{icml2024}

\usepackage{amsmath}
\usepackage{amssymb}
\usepackage{mathtools}
\usepackage{amsthm}
\usepackage{makecell}
\usepackage{bm}
\usepackage{bbm}
\usepackage{algorithm}
\usepackage{algorithmic}

\usepackage[capitalize,noabbrev]{cleveref}

\theoremstyle{plain}
\newtheorem{theorem}{Theorem}[section]
\newtheorem{proposition}[theorem]{Proposition}
\newtheorem{lemma}[theorem]{Lemma}

\theoremstyle{definition}

\theoremstyle{remark}
\newtheorem{remark}[theorem]{Remark}

\usepackage[textsize=tiny]{todonotes}

\newcommand{\bx}[1]{\mathbf{x}_{#1}}
\newcommand{\scstate}[1]{\bm{\eta}_{#1}}

\newcommand{\by}{\mathbf{y}}
\newcommand{\mask}{\mathbf{M}}
\newcommand{\bff}{\mathbf{f}}
\newcommand{\bw}{\mathbf{w}}
\newcommand{\bu}[1]{\mathbf{u}_{#1}}
\newcommand{\dd}{\textrm{d}}
\newcommand{\tf}{\tilde{\mathbf{f}}}
\newcommand{\terminalonly}{\phi}
\newcommand{\mQ}{\mathcal{Q}}
\newcommand{\initial}{\bar{\scstate{}}_0}
\newcommand{\diracp}{p_{\bar{\bm{\eta}}_0}}
\newcommand{\terminal}[1]{\phi(#1)}
\newcommand{\defeq}{\stackrel{\Delta}{=}}

\usepackage{xspace}
\newcommand{\oursfull}{Stochastic Control Guidance\xspace}
\newcommand{\ours}{SCG\xspace}

\DeclareMathOperator*{\argmax}{argmax}

\icmltitlerunning{Symbolic Music Generation with Non-Differentiable Rule Guided  Diffusion}

\begin{document}

\twocolumn[
\icmltitle{Symbolic Music Generation with Non-Differentiable Rule Guided Diffusion}



\icmlsetsymbol{equal}{*}

\begin{icmlauthorlist}
\icmlauthor{Yujia Huang}{caltech}
\icmlauthor{Adishree Ghatare}{caltech}
\icmlauthor{Yuanzhe Liu}{rpi}
\icmlauthor{Ziniu Hu}{caltech}
\icmlauthor{Qinsheng Zhang}{nvidia}
\icmlauthor{Chandramouli S Sastry}{dal,vector}
\icmlauthor{Siddharth Gururani}{nvidia}
\icmlauthor{Sageev Oore}{dal,vector}
\icmlauthor{Yisong Yue}{caltech}
\end{icmlauthorlist}

\icmlaffiliation{caltech}{California Institute of Technology}
\icmlaffiliation{rpi}{Rensselaer Polytechnic Institute}
\icmlaffiliation{nvidia}{NVIDIA}
\icmlaffiliation{dal}{Dalhousie University}
\icmlaffiliation{vector}{Vector Institute}

\icmlcorrespondingauthor{Yujia Huang}{yjhuang@caltech.edu}

\icmlkeywords{Machine Learning, ICML}

\vskip 0.3in
]



\printAffiliationsAndNotice{}  

\begin{abstract}
We study the problem of symbolic music generation (e.g., generating piano rolls), with a technical focus on non-differentiable rule guidance.    
Musical rules are often expressed in symbolic form on note characteristics, such as note density or chord progression, many of which are non-differentiable which pose a challenge when using them for guided diffusion.
We propose \oursfull (\ours), a novel guidance method that only requires forward evaluation of rule functions that can work with pre-trained diffusion models in a plug-and-play way, thus achieving training-free guidance for non-differentiable rules for the first time.  
Additionally, we introduce a latent diffusion architecture for symbolic music generation with high time resolution, which can be composed with SCG in a plug-and-play fashion. Compared to standard strong baselines in symbolic music generation, this framework demonstrates marked advancements in music quality and rule-based controllability, outperforming current state-of-the-art generators in a variety of settings. For detailed demonstrations, code and model checkpoints, please visit our \href{https://scg-rule-guided-music.github.io/}{project website}.
\end{abstract}

\section{Introduction}
We are interested in developing methods for controllable symbolic music generation.  There has been rapid progress in the development of modern generative models for symbolic music~\citep{huang2018music, huang2020pop, hsiao2021compound, min2023polyffusion}.  
To facilitate interaction between human composers and these models, it is crucial for these models to adhere to specific musical rules, such as chord progression, during the composition process.  

A common method to incorporate rules in generative models is to train with rule labels \cite{choi2020encoding, wu2023musemorphose, von2022figaro}.  However, integrating multiple musical rules during the training phase poses a significant challenge. Continuously updating model parameters to accommodate each new rule is not only costly but also will soon become impractical for compositions that involve many rules. Hence, there's a growing need for a method to guide pre-trained generative models in generating samples that conform to specific rules in a more flexible, light-weight, or plug-and-play manner.

Diffusion models \cite{ho2020denoising,song2020score} have emerged as a powerful generative modeling approach in many domains including images \cite{dhariwal2021diffusion}, audio \cite{huang2023noise2music} and video \cite{ho2022imagen}. A key feature of diffusion models is that they allow for post-hoc guidance of pre-trained models. Recent works have demonstrated success in guiding diffusion models with differentiable losses in a plug-and-play manner \cite{chung2022diffusion, song2023loss}. Starting from Gaussian noise, diffusion models generate samples from coarse to fine. The key idea of guidance is to update each intermediate step with the gradient of the loss.
However, there are still two challenges to generate symbolic music with rule guidance:
First, many rules (e.g. note density) are not differentiable. 
Second, they may be black box APIs that hinder backpropagation. 

\begin{figure}
\centering
    	\includegraphics[width=\linewidth]{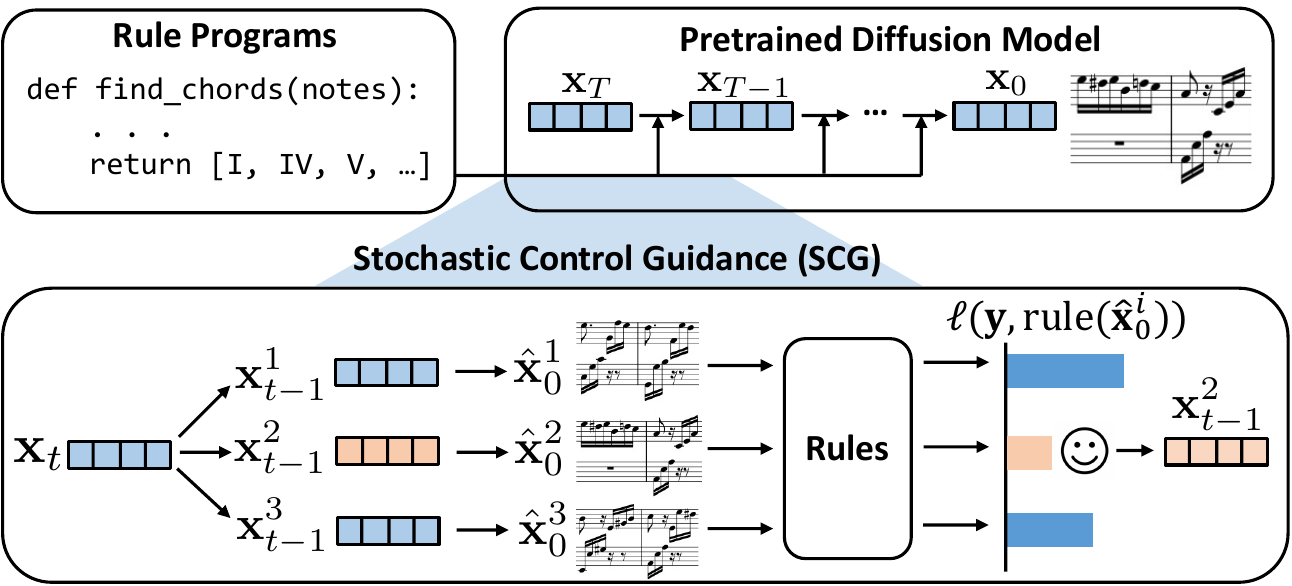}
        \vspace{-0.25in}
        \caption{Overview of Stochastic Control Guidance (SCG) for plug-and-play non-differentiable rule guided generation. At each sampling step, we sample several realizations of the next step, and select the one yielding the most rule-compliant clean sample.}
\label{fig:overview}
\vspace{-0.2in}
\end{figure}

To this end, we propose \oursfull~(\ours), a new algorithm that enables plug-and-play guidance in diffusion models for non-differentiable rules. Our algorithm is inspired by stochastic control, where we pose the problem of generating samples that follow rule guidance as optimal control within a stochastic dynamical system. We obtain the analytical form of optimal control via path integral control theory~\cite {theodorou2010generalized}, and adapt it to an efficient implementation within diffusion models. Specifically, we generate multiple realizations at each sampling step, and select the one that best follows the target (Figure \ref{fig:overview}). This process only requires forward evaluation of rule functions, making it applicable to non-differentiable rules. 

To develop a practical overall framework, we also introduce a latent diffusion architecture with a transformer backbone for symbolic music generation. This architecture is able to generate dynamic music performances at 10ms time resolution, which is a significant challenge for standard pixel space diffusion models.

Our framework demonstrates state-of-the-art performance in various music generation tasks, offering superior rule guidance over popular methods and enabling musicians to effectively use it as a compositional tool.
Our code is available \href{https://github.com/yjhuangcd/rule-guided-music}{here}.

In summary, our contributions are as follows:
\vspace{-0.1in}
\begin{itemize}[itemsep=0.5mm]
    \item We introduce \oursfull~(\ours), which achieves plug-and-play guidance in diffusion models for non-differentiable rules.
    \item We provide a theoretical justification of \ours from a stochastic control perspective.
    \item We introduce a latent diffusion model architecture for symbolic music generation with high time resolution.
    \item We demonstrate that our framework enables flexible, interpretable and controllable symbolic music generation in a variety of tasks. 
\end{itemize}

\section{Related Works}
Current symbolic music generation methods are mainly divided into MIDI token-based and piano roll-based approaches. MIDI-based methods treat music as sequences of discrete tokens, often using transformers for MIDI token generation \cite{huang2018music,huang2020pop,ren2020popmag,hsiao2021compound}. Piano roll representations, resembling image formats with time on the horizontal axis and pitches vertically, have inspired the use of image generative models like GANs \cite{yang2017midinet, dong2018musegan} for their generation. Recent efforts \cite{atassi2023generating, min2023polyffusion} apply diffusion models to generate binary, quantized piano rolls. Our work extends this by incorporating velocity and pedal information into piano rolls and employing a finer time resolution of 10 ms, thereby facilitating the generation of more dynamic piano performances.

Another line of research seeks to enhance control over certain attributes in the generated music. Some studies \cite{brunner2018midi,roberts2018hierarchical} have leveraged VAE models to learn a disentangled latent space, achieving controllability over specific attributes by manipulating latents in designated directions. Further, various works have conditioned LSTMs \cite{meade2019exploring} or transformers on different factors like style \cite{choi2020encoding}, note density \cite{wu2023musemorphose}, or attributes like time signature, instruments, and chords \cite{von2022figaro}. However, these methods are limited to predefined attributes and are not easily extendable to new attributes due to the necessity of conditioning on labels during training.

Recent developments in the use of diffusion models for symbolic music generation have adapted controllable image generation techniques. Examples include generating complementary parts given melody/accompaniment (inpainting), bridging two music segments (infilling) \cite{min2023polyffusion}, extending existing music pieces (outpainting), and generating piano rolls from stroke piano rolls \cite{zhang2023sdmuse}. Yet, when it comes to rule-based guidance, existing approaches still require training on specific attributes, such as chord progression \cite{min2023polyffusion,li2023melodydiffusion}, limiting their adaptability for composers desiring to incorporate new rules. 
Our work enables flexible rule-based guidance via \ours. Additionally, our method is compatible with other diffusion model techniques like inpainting, outpainting, and editing, further enhancing its versatility in music generation.

The conceptualization of stochastic optimal control in diffusion models has spurred theoretical advancements and practical applications. Path integral theory \cite{kappen2005path} provides an efficient way of solving stochastic optimal control problems. 
\citet{zhang2021path} employed it to transform a simple Ornstein–Uhlenbeck process to a novel process whose target distribution matches given marginal distribution.
Further extending this framework, ~\citet{berner2022optimal,vargas2023denoising} established a novel link between stochastic optimal control problems and generative models, interconnected through stochastic differential equations.

\section{Background}
\textbf{Score-based diffusion models.}
Diffusion models generate data by reversing a diffusion process. Let $p(\bx{})$ be the unknown data distribution, the forward diffusion process $\{\bx{t}\}_{t \in [0,T]}$ diffuse $p(\bx{})$ to a noise distribution that is easy to sample from (e.g. standard Gaussian distribution). \citet{song2020score} models the forward diffusion process as the solution to an SDE:
\begin{equation}
    \dd \bx{} = \bff(\bx{}, t) \dd t + g(t) \dd \bw,
\end{equation}
where the initial condition $\bx{0} := \bx{} \sim p(\bx{})$, $\bff: \mathbb{R}^d \times \mathbb{R} \rightarrow \mathbb{R}^d$ is the drift coefficient, $g: \mathbb{R} \rightarrow \mathbb{R}$ is the diffusion coefficient and $\bw \in \mathbb{R}^d$ is a standard Wiener process. 

Let $p_t(\bx{})$ denote the marginal distribution of $\bx{t}$. The diffusion and drift coefficient can be properly designed such that $p_T(\bx) \approx \mathcal{N}(\mathbf{0}, \mathbf{I}_d)$. In this paper, we consider the VP-SDE \cite{song2020score}, where $\bff(\bx{}, t) := -\frac{1}{2}\beta(t)\bx{}$ and $g(t):=\sqrt{\beta(t)}$, where $\beta(t)$ is a noise schedule. DDPM \cite{ho2020denoising} can be regarded as a discretization of VP-SDE.

Samples are generated using the reverse-time SDE:
\begin{equation}
    \dd \bx{t} = \left[ \bff(\bx{t}, t) - g(t)^2 \nabla_{\bx{t}} \log p_t(\bx{t}) \right] \dd t + g(t) \dd \bar{\mathbf{w}}_t, 
    \label{eq:reverse-sde}
\end{equation}
where $\bff(\bx{t}, t): \mathbb{R}^d \rightarrow \mathbb{R}$ is the drift coefficient, $g: \mathbb{R} \rightarrow \mathbb{R}$ is the diffusion coefficient, $\dd t$ is an infinitesimal negative time step and $\bar{\mathbf{w}}_t$ is a standard reverse-time Wiener process. Sampling $\bx{T} \sim p_T(\bx{}) = \mathcal{N}(\mathbf{0}, \mathbf{I})$ and solving the above SDE from $t=T$ to $t=0$ produces samples from the data distribution: $\bx{0} \sim p_0(\bx{}) = p(\bx{})$.

Since the data distribution is unknown, it is popular to approximate the score function $\nabla_{\bx{t}} \log p_t(\bx{t})$ via a neural network $\mathbf{s}_{\theta}(\bx{},t)$ and train it with a weighted sum of denoising score matching objectives \cite{song2020score}. 

\textbf{Classifier and Classifier-free Guidance.}
Guided diffusion models generates samples from $p(\bx{}|\by)$ given label $\by$. Classifier guidance \cite{dhariwal2021diffusion} achieves this by training a classifier $p_t(\by | \bx{t})$ on the noisy sample and label pair, and mix its gradient with the score of the diffusion model during sampling. The conditional score function becomes $\nabla_{\bx{t}} \log p_t(\bx{t}) + \omega \nabla_{\bx{t}} \log p_t(\by | \bx{t})$, where $\omega$ is called guidance scale. This approximates the samples from the distribution $\tilde{p}(\bx{t} | \by) \propto p(\bx{t}) p (\by | \bx{t})^\omega$. Classifier guidance is able to guide a pre-trained generative model at the cost of training an extra classifier on the noisy data. 

Classifier-free guidance \cite{ho2022classifier} avoids training classifiers by jointly training conditional and unconditional diffusion models, and combining their score estimates during sampling. The mixed score function becomes $(1+\omega) \nabla_{\bx{t}} \log p_t(\bx{t} | \by) - \omega \nabla_{\bx{t}} \log p_t(\bx{t})$, where $\omega$ is the guidance strength. Despite easy implementation, it is expensive to extend classifier-free guidance to unknown or composite labels, because it requires re-training the diffusion model.

\textbf{Loss-Guided Diffusion.} To reduce the need of additional training for conditional generation, methods have been proposed to guided diffusion models to generate samples in a plug-and-play way. Instead of training a classifier to approximate $p (\by | \bx{t})$, Diffusion Posterior Sampling (DPS) \cite{chung2022diffusion} uses $p(\by | \hat{\bx{}}_0)$, where $\hat{\bx{}}_0 := \mathbb{E}[\bx{0} | \bx{t}]$ is obtained through the Tweedie's formula \cite{efron2011tweedie}:
\begin{equation}
\label{eq:tweedie}
    \hat{\bx{}}_0 = \frac{1}{\sqrt{\bar{\alpha}(t)}}(\bx{t} + (1 - \bar{\alpha}(t)) \nabla_{\bx{t}} \log p_t(\bx{t})).
\end{equation}
Recall that $p(\by | \bx{t})$ can be factorized as: 
$$p(\by | \bx{t}) = \int p(\by | \bx{0}) p(\bx{0} | \bx{t}) \dd \bx{0} = \mathbb{E}_{\bx{0} \sim p(\bx{0} | \bx{t})} p (\by | \bx{0}).$$ 
DPS uses a point estimation of this quantity. Later work \cite{song2023loss} proposes to use Monte-Carlo estimation of this by sampling from approximated $p(\bx{0} | \bx{t})$. However, these methods requires the loss function used to specify the condition to be differentiable. Many symbolic rules we consider in this paper are non-differentiable.

\section{Non-Differentiable Rule Guidance}
\label{sec:rule-guided-diffusion}
We now present \oursfull for non-differentiable rule guidance in diffusion models.  We start with defining rule guidance in Section \ref{sec:rule_guide_def}. Inspired by stochastic control (Section \ref{sec:stochastic_control_formulation}), we define a value function as a loss measuring (lack of) rule adherence, and show that optimal control steers the reverse diffusion to the target distribution. We then discuss practical algorithms (Section \ref{sec:practical_algo}).  We conclude by establishing a general theoretical connection that enables many guidance methods to be viewed through the lens of stochastic optimal control (Section \ref{sec:general_sc}).

\subsection{Rule Guidance Problem}
\label{sec:rule_guide_def}
Assume that we have a pre-trained diffusion model that can sample from the data distribution $p(\bx{})$, and a loss function $\ell_{\by}: \mathcal{X} \rightarrow \mathbb{R}$ that characterizes how well a sample follows some conditions $\by$: $p(\by|\bx{}) \propto e^{-\ell_{\by}(\bx{})}$. Our goal is to sample from the following distribution:
\begin{equation}
    p(\bx{}|\by) = p(\bx{}) \frac{e^{-\ell_{\by}(\bx{})}}{Z} \propto p(\bx{}) p(\by | \bx{}),\label{eq:posterior}
\end{equation}
where $Z = \int_{\bx{}} p(\bx{}) e^{-\ell_{\by}(\bx{})} \dd \bx{} $. 

A central challenge that we tackle is that many musical rules  are non-differentiable, which makes sampling from Eq.~\ref{eq:posterior} difficult. For instance, let $\bx{}=[x_1, x_2, ..., x_n] \in [0,1]^n$ be a vector where each $x_i$ represents the volume of a note, so that the note density is computed as $\operatorname{ND}(\bx{})=\sum_{i=1}^n \mathbbm{1}(x_i > \epsilon)$, where $\epsilon$ is a small number. Let $\by=y$ be a scalar that represents the target note density. Then the loss is defined as $\ell_{\by}(\bx{}) = |y - \operatorname{ND}(\bx{})|$, which is non-differentiable. 

\subsection{Guidance via Stochastic Control}
\label{sec:stochastic_control_formulation}
The pre-trained diffusion model generates samples using the reverse-time SDE (Eq.~\ref{eq:reverse-sde}).
Let $\scstate{t} = \bx{T-t}$, and $\tf(\scstate{t},t)=\bff(\scstate{t}, t) - g(t)^2 \nabla_{\scstate{t}} \log p_t(\scstate{t})$. We can rewrite Eq.~\ref{eq:reverse-sde}  as: 
\begin{equation}
\label{eq:uncontrolled_sde}
    \textrm{d} \scstate{t} = \tf(\scstate{t},t) \dd t + g(t) \dd \bw_t, 
\end{equation}
where $\dd t$ is an infinitesimal time step and $\dd \bw$ is a standard Wiener process. Sampling $\scstate{0} \sim \mathcal{N}(\mathbf{0}, \mathbf{I})$ and solving the above SDE from $t=0$ to $t=T$ produces samples from the data distribution. 

We want to find a control $\bu{}(\scstate{t},t)$, such that solving the following SDE yields samples from target distribution $p(\scstate{}|\by)$:
\begin{equation}
\label{eq:controlled_sde}
    \textrm{d} \scstate{t} = \tf(\scstate{t},t) \dd t + g(t) (\bu{}(\scstate{t},t) \dd t + \dd \bw_t). 
\end{equation}
We use $\bu{t} := \bu{}(\scstate{t},t)$ and $\tf_t := \tf(\scstate{t},t)$ for brevity, noting they are state-dependent.

Considering the stochastic dynamical system in Eq.~\ref{eq:controlled_sde} for $0 \leq t \leq T$ and initial state $\scstate{0} = \initial$, we address the optimal control problem associated with the cost function $C_u(\scstate{t},t)$, which is defined as the expectation over all stochastic trajectories starting at $\scstate{t}$ with control function $\bu{t}$:
\begin{equation}
\label{eq:cost}
    C_u(\scstate{t},t) = \mathbb{E} \left[ \terminal{\scstate{T}} + \int_t^T \frac{1}{2} \| \bu{t} \|^2 \dd t \right].
\end{equation}
It is known that the optimal control policy admits an analytical solution~\citep{pavon1989stochastic}: 
\begin{equation}
    \label{eq:optimal_control}
    \bu{t}^* = - g(t) \nabla_{\scstate{}} V(\scstate{}, t),
\end{equation}
where function $V(\scstate{}, t)$, known as the \emph{value} function, is the solution to celebrated stochastic Hamilton-Jacobi-Bellman (HJB) equation~\cite{evans2022partial}:
\begin{align}
    -\partial_t V(\scstate{}, t) &= - \frac{1}{2} g(t)^2
    (\nabla_{\scstate{}} V)^\top (\nabla_{\scstate{}} V) \nonumber \\
    &+ (\nabla_{\scstate{}} V)^\top \tf_t + \frac{1}{2} g(t)^2 \textrm{Tr}(\nabla_{\scstate{}\scstate{}}^2 V),
    \label{eq:HJB}
\end{align}
with boundary condition $V(\scstate{}, T) = \terminal{\scstate{}}$.

\textbf{Path Integral Control.}
Although solving HJB in Eq.~\ref{eq:HJB} is nontrivial due to its non-linearity w.r.t.~$V$, using an exponential transformation $\Psi(\scstate{}, t) = e^{-V(\scstate{}, t)}$ yields a linear HJB equation in $\Psi$:
\begin{equation}
\label{eq:linear_hjb}
    -\partial_t \Psi(\scstate{},t) = \left( \tf_t ^\top \nabla_{\scstate{}} + \frac{1}{2} g(t)^2 \textrm{Tr}(\nabla_{\scstate{}\scstate{}}^2) \right) \Psi(\scstate{},t),
\end{equation}
with boundary condition $\Psi(\scstate{}, T) = e^{-\terminal{\scstate{}}}$. 
We call $\Psi$ the \emph{desirability} function as it is inversely related to the value $V$.

Let $\Omega = C([0, T]; \mathbb{R}^d)$ be the space consisting of all possible continuous-time stochastic trajectories $\tau = \{ \scstate{t}, 0 \leq t \leq T \}$, and $\mQ^0$ be the measure induced by an uncontrolled stochastic process (Eq.~\ref{eq:uncontrolled_sde}).
Then the linear HJB equation has the following solution according to the Feynman-Kac formula \cite{oksendal2003stochastic}:
\begin{equation}
\label{eq:feynman-kac}
    \Psi(\scstate{}, t) = \mathbb{E}_{\mQ^0} \left[ e^{-\terminal{\scstate{T}}} | \scstate{t}=\scstate{} \right].
\end{equation}
Eq.~\ref{eq:feynman-kac} shows that the value function can be computed by \emph{only forward sampling the uncontrolled process} without knowing the optimal control policy. Plugging Eq~\ref{eq:feynman-kac} into Eq~\ref{eq:optimal_control} yields the analytic optimal policy, which aligns with the well-known path integral control approach~\citep{theodorou2010generalized,theodorou2015nonlinear,fleming1982optimal}:
\begin{align}
    \label{eq:optimal_u_grad}
    \bu{t}^*(\scstate{}) \dd t &= g(t) \nabla_{\scstate{}}\log \Psi(\scstate{},t) \dd t \\
    \label{eq:optimal_u_pi}
    &= \frac{\mathbb{E}_{\mQ^0}\left[ e^{-\terminal{\scstate{T}}} \dd \bw_t | \scstate{t} = \scstate{} \right]}{\mathbb{E}_{\mQ^0} \left[ e^{- \terminal{\scstate{T}}} | \scstate{t} = \scstate{} \right]}.
\end{align}

Next, we show that using the above optimal control, we can guide the generation process to produce samples from the target conditional distribution $p(\scstate{} | \by)$.
\begin{theorem}[proof in Appendix \ref{sec:thm1_proof}]
\label{thm:path_int_guidance}
    Consider the dynamical system in Eq.~\ref{eq:controlled_sde}. For a terminal cost defined as $\terminal{\scstate{T}} \defeq \ell_y(\scstate{T}) \defeq - \log p(\by | \scstate{T}) + \mathtt{const}$, and initial condition $\scstate{0} \sim \mathcal{N}(\mathbf{0}, \mathbf{I})$, the terminal distribution induced by the optimal control policy $\bu{t}^*$ (Eq.~\ref{eq:optimal_u_pi}) is:
    \begin{equation}
        \mQ^*(\scstate{T}) = p(\scstate{T}|\by).
    \end{equation}
\end{theorem}

\begin{algorithm}[t]
   \caption{Stochastic Control Guided DDPM sampling}
   \label{alg:sampling_algo}
\begin{algorithmic}
    \STATE \textbf{Require:} Loss function $\ell_y$, rule target $\by$, number of samples $n$, forward process variances $\beta_t$, $\alpha_t := 1-\beta_t$, $\bar{\alpha}_t := \prod_{s=1}^{t} \alpha_s$.
   \STATE $\bx{T} \sim \mathcal{N}(\mathbf{0}, \mathbf{I})$
   \FOR{$t=T$ {\bfseries to} $1$}
        \STATE $\triangleright$ Compute the posterior mean of $\bx{t-1}$.
        \STATE $\hat{\bx{}}_{t-1} = \frac{1}{\sqrt{\alpha}_t}\left(\mathbf{x}_t - \frac{1-\alpha_t}{\sqrt{1-\bar{\alpha}_t}}\epsilon_\theta\left(\mathbf{x}_t, t\right)\right)$
        \IF{$t>1$}
            \STATE $\triangleright$ Sampling possible next steps.
            \STATE $\mathbf{x}_{t-1}^i = \hat{\bx{}}_{t-1} + \sigma_t \mathbf{z}^i$, with $\mathbf{z}^1, ..., \mathbf{z}^n \sim \mathcal{N}(\mathbf{0}, \mathbf{I})$
            \STATE $\triangleright$ Estimate the clean sample from noisy sample.
            \STATE $\hat{\mathbf{x}}_0^i = \frac{1}{\sqrt{\bar{\alpha}_{t-1}}}\left(\mathbf{x}_{t-1}^i - \sqrt{1-\bar{\alpha}_{t-1}}\epsilon_\theta\left(\mathbf{x}^i_{t-1}, t-1\right)\right)$
            \STATE $\triangleright$ Find the direction that minimizes the loss.
            \STATE $k = \argmax_i \log p(y | \hat{\bx{}}_0^i ) = \argmax_i -\ell_y(\hat{\bx{}}_0^i)$
            \STATE $\mathbf{x}_{t-1} = \mathbf{x}_{t-1}^k$
        \ELSE
            \STATE $\mathbf{x}_{t-1} = \hat{\bx{}}_{t-1}$
        \ENDIF
   \ENDFOR
   \STATE {\bfseries return: } $\bx{0}$
\end{algorithmic}
\end{algorithm}
\vspace{-0.15in}

\subsection{Practical Algorithms}
\label{sec:practical_algo}
\textbf{Approximation of the Optimal Control.}
In practice, it is expensive to compute Eq.~\ref{eq:optimal_u_pi}, because one needs to unroll the whole trajectory to get $\scstate{T}$. 
Instead of using Eq.~\ref{eq:optimal_u_pi} as our optimal control, we set $\bu{t} \dd t + \dd \bw$ to the following:
\begin{equation}
\label{eq:argmax_approx}
    \argmax_{\dd \bw_t} -\ell_y(\hat{\scstate{}}_T),
\end{equation}
where $\hat{\scstate{}}_T = \mathbb{E} \left[ \scstate{T} | \scstate{t+\dd t} \right]$ can be obtained via Tweedie's Formula (Eq.~\ref{eq:tweedie}), which is a one-step computation and much cheaper than solving the whole trajectory.

Eq.~\ref{eq:argmax_approx} is an approximation to a tempered version of Eq.~\ref{eq:optimal_u_pi}. 
Consider the terminal cost is defined with a scaling factor $K$, i.e. $\terminal{\scstate{T}} = \ell_y(\scstate{T}) / K$. 
When $K \rightarrow 0$, Eq.~\ref{eq:optimal_u_pi} becomes:
\begin{equation}
\label{eq:argmax_optimal_u}
    \argmax_{\dd \bw_t} \max_{\tau} -\ell_y(\scstate{T} | \scstate{t+\dd t}),
\end{equation}
where $\scstate{t+\dd t} = \scstate{t} + \tf(\scstate{t},t) \dd t + g(t) \dd \bw_t$, and $\tau : [t + \dd t, T] \rightarrow \mathbb{R}^d$ represents a trajectory. 
The solution of Eq.~\ref{eq:argmax_approx} optimizes a lower bound of the objective in Eq.~\ref{eq:argmax_optimal_u}:
\begin{equation}
    \max_{\dd \bw_t, \tau} -\ell_y(\scstate{T} | \scstate{t+\dd t}) \geq \max_{\dd \bw_t} -\ell_y(\mathbb{E} \left[ \scstate{T} | \scstate{t+\dd t} \right]).
\end{equation}

\textbf{Intuition.} 
Our \ours algorithm implemented with DDPM sampling \cite{ho2020denoising} is outlined in Algorithm \ref{alg:sampling_algo} and illustrated in Figure \ref{fig:overview}, where we use $\bx{t} \defeq \scstate{T-t}$ to denote the intermediate states following conventions of diffusion model notations. The intuition is that we select the direction that leads to the most probable sample at each step. For every step $t$ in the sampling process, given $\bx{t}$, we compute multiple realizations of the next step $\bx{t-1}$, estimate the corresponding clean sample $\hat{\mathbf{x}}_0$, and choose the $\bx{t-1}$ that leads to the lowest loss $\ell_{\by}(\hat{\mathbf{x}}_0)$. Notably, we only need to evaluate the forward pass of the rule function, and there is no need to evaluate or estimate its gradient, making our method suitable for non-differentiable and black-box rule functions.
Furthermore, it is also compatible with other stochastic sampling procedure in diffusion models (Appendix \ref{sec:appen_algo}).

\subsection{General Theoretical Connection} 
\label{sec:general_sc}

In this section, we show a general connection (Proposition \ref{thm:desirability_func}) that enables many guidance methods to be viewed through the lens of stochastic optimal control.

\begin{proposition}[proof in Appendix \ref{sec:thm2_proof}]
\label{thm:desirability_func}
Consider the dynamical system in Eq.~\ref{eq:controlled_sde} with terminal cost $\terminal{\scstate{T}} \defeq - \log p(\by | \scstate{T}) + \mathtt{const}$. We have: $\Psi(\scstate{t}, t) = c \cdot p(\by | \scstate{t})$.
\end{proposition}
Proposition \ref{thm:desirability_func} says that the desirability function equals to the likelihood function. Then many popular guidance techniques can be seen as different implementations of the optimal control  following Eq.~\ref{eq:optimal_u_grad}):
\begin{equation*}
    g(t) \nabla_{\scstate{t}}\log p(\by | \scstate{t}) = g(t) \nabla_{\scstate{t}}\log \Psi(\scstate{t},t) = \bu{t}^*(\scstate{t}).
\end{equation*}
Classifier guidance \cite{dhariwal2021diffusion} trains a neural network on noisy data pair $\{ \scstate{t}, \by \}$ to approximate $\Psi(\scstate{t},t)$, and differentiate through it to obtain $\bu{t}^*(\scstate{})$. 

DPS \cite{chung2022diffusion} avoids training a surrogate model by approximating $\Psi(\scstate{t},t)$ with $\Psi(\hat{\scstate{}}_T,T)$, where $\hat{\scstate{}}_T$ is the posterior mean that can be obtained through the Tweedie's formula (Eq.~\ref{eq:tweedie}).
Since $\nabla_{\scstate{t}} \Psi(\hat{\scstate{}}_T,T) = \frac{\partial \Psi(\hat{\scstate{}}_T,T)}{\partial \hat{\scstate{}}_T} \frac{\partial \hat{\scstate{}}_T}{\partial \scstate{t}}$, it requires $\Psi(\hat{\scstate{}}_T,T) \propto e^{-\ell_{\by}(\hat{\scstate{}}_T)}$ to be differentiable.

In contrast, our approach is inspired by path integral control, and only needs the forward evaluation of the rule function (Eq.~\ref{eq:argmax_approx}). Therefore, our method does not require the rule function to be differentiable.

\section{Latent Diffusion Architecture}
To arrive at a practical overall framework, we develop a latent diffusion architecture tailored towards symbolic music generation, and in particular able to generate at 10ms time resolution.  This architecture can be combined with Stochastic Control Guidance in a plug-and-play fashion.

\textbf{Data Representation.}
We represent symbolic music as a 3-channel tensor.
Each column in this representation accounts for a 10 ms timeframe. The first channel is the piano roll, where horizontal axis represents time and vertical axis represents pitch. Each element takes value from 0-127, indicating the velocity (volume) of the note. The second channel is the onset roll, consisting of binary values that denote the presence of note onsets. The third channel is the pedal roll, representing the sustain pedal control for each timeframe. 

\begin{figure}[h]
\centering
\hspace*{0.4cm}
    	\includegraphics[width=0.995\linewidth]{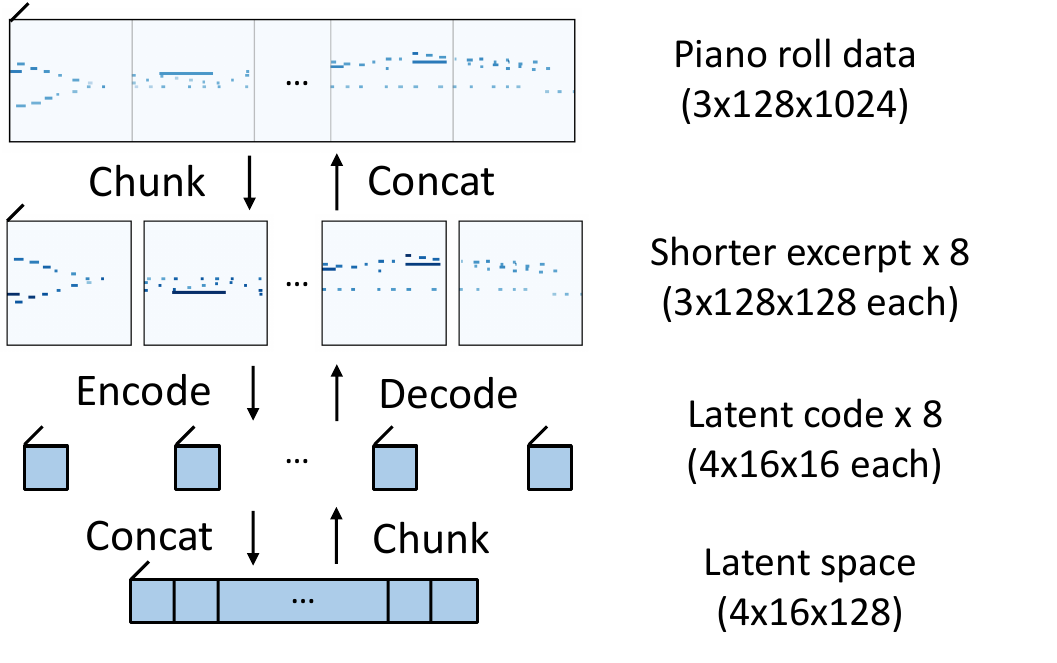}
     \vspace{-0.2in}
        \caption{We use a VAE to encode piano roll segments to latent space and concatenate them for the next stage of diffusion training.}
\label{fig:vae}
\end{figure}

\textbf{Model architecture.}
We first use a VAE model to encode short segments of piano rolls of shape $3 \times 128 \times 128$ into a latent space. Then we concatenate the latent codes and train a diffusion model to capture their joint distribution (Figure \ref{fig:vae}). 
For the VAE, we use the same architecture following \cite{rombach2022high}. 
The training involves a denoising objective in conjunction with KL regularization: we introduce musically semantic perturbations (such as adding adjacent notes) to the data and train the model to revert to the original, unperturbed data. Both KL regularization and the denoising objective have proven indispensable for developing diffusion models with robust generative capabilities in subsequent stages.

For the diffusion model, we use the DiT architecture \cite{peebles2023scalable}. In contrast to the standard U-Net, the transformer backbone is more adept at handling sequences of latent tokens. Rather than absolute position encoding, we use rotary position embedding \cite{su2023roformer} to better generalize across various input lengths. 
We train the diffusion model on piano rolls of length 1024 (10.24 s). To generate musical excerpts of arbitrary length, we apply DiffCollage \cite{zhang2023diffcollage} to aggregate the score function of shorter music segments.

\section{Experiments}
We evaluate our method on a wide range of symbolic music generation tasks: unconditional generation (Sec \ref{sec:uncond_gen}), individual rule guidance (Sec \ref{sec:single_rule}), composite rule guidance (Sec \ref{sec:multiple_rule}) and editing (Appendix \ref{sec:edit}). We perform ablation studies in Sec \ref{sec:abla} and subjective evaluation in Sec \ref{sec:subj_test}. In addition, we demonstrate that our method can be used as a compositional tool for musicians in Sec \ref{sec:compositionExamples}.

\subsection{Experimental Settings} 
\label{sec:exp_setting}
\textbf{Data.}
We train our model on several piano midi datasets that cover both classical and pop genres.  The MAESTRO dataset \cite{hawthorne2018enabling} has about 1200 pieces of classical piano performances with expressive dynamics, resulting in about 200 hours of music performance. In addition, we crawled about 14k MIDI files from Muscore in classical, religion, and soundtrack genres across all skill levels, yielding about 700 hours of data.
We also used two Pop piano datasets: Pop1k7 \cite{hsiao2021compound} and Pop909 \cite{wang2020pop909} that contain 108 hours and 60 hours of pop piano midi files.

\textbf{Training and Inference Setup.}
We first train a VAE model to encode piano rolls to latent space, then fix the VAE and train a diffusion model on this space.
The diffusion model is trained with dataset-based conditioning: classical performance (Maestro), classical sheet music (Muscore) and Pop (pop1k7 and pop909), following Classifier-free Guidance \cite{ho2022classifier} with a dropout rate of $0.1$. We train the model for 1.2M steps and use DDPM \cite{ho2020denoising} with 1000 steps as the default sampling method unless stated otherwise. All experiments are run on NVIDIA A100-SXM4 GPUs.

\subsection{Unconditional Generation}
\label{sec:uncond_gen}
\textbf{Baselines.}
We compare with state-of-the-art symbolic music generators trained on various datasets (Table \ref{tab:music_baseline}). 
\begin{table}[h]
    \centering
    \resizebox{\columnwidth}{!}{
    \begin{tabular}{lccc}
    \toprule
       Method & Model & Dataset & Representation \\
    \midrule
        MusicTr \cite{huang2018music} & Transformer & Maestro & MIDI-like \\
        Remi \cite{huang2020pop} & Transformer & Pop775 & REMI \\
        CPW \cite{hsiao2021compound} & Transformer & Pop1k7 & CP \\
        PolyDiff \cite{min2023polyffusion} & Diffusion & POP909 & Piano roll \\
    \bottomrule
    \end{tabular}
    }
    \vspace{-0.1in}
    \caption{Baselines for unconditional music generation.}
    \label{tab:music_baseline}
    \vspace{-0.1in}
\end{table}

\textbf{Objective Metrics}. It is worth mentioning that quantitative evaluation of music quality remains an open problem \cite{yin2023deep}. Nevertheless, we use the average overlapping area (OA) between the intra-set and inter-set distribution of 7 musical attributes (pitch range, note density, etc.) proposed in \cite{yang2020evaluation} as the objective metric for music quality.
As a sanity check, we compare a subset of the training dataset with another subset (denoted by GT in Table \ref{tab:uncond_gen}), and find that GT on all the datasets achieves the highest average OA. This indicates that this metric is a reasonable necessary condition for good generated music quality.

\textbf{Results.} The evaluation results are in Table \ref{tab:uncond_gen}, highlighting the highest values (excluding GT) in bold. Our method achieves the highest average OA on all the datasets. The detailed results for all 7 OA metrics are in Table \ref{tab:detail_uncond_gen}, Appendix \ref{sec:appen_additional_exp}.
The baselines are trained on individual dataset, and do not generalize well across datasets. MusicTr has the second-best overall rating for classical music (Maestro and Muscore), while it holds the lowest rating for pop music. CPW, on the other hand, ranks second in pop music but has the lowest rating in classical music.
In contrast, our model delivers strong performance consistently across all the datasets.

\begin{table*}[h]
\centering
\resizebox{0.9\textwidth}{!}{
\begin{tabular}{lcccccc}
\toprule
Dataset       & GT     & MusicTr     & Remi      & CPW     & PolyDiff      & \textbf{\ours (ours)}           \\
\midrule
Maestro  & $0.944 \pm 0.002$  &  $0.903 \pm 0.005$  &  $0.847 \pm 0.005$  &  $0.801 \pm 0.006$  &  $0.842 \pm 0.007$  &  $\mathbf{0.943 \pm 0.003}$  \\
Muscore  &  $0.945 \pm 0.004$  &  $0.901 \pm 0.004$  &  $0.879 \pm 0.006$  &  $0.843 \pm 0.007$  &  $0.845 \pm 0.004$  &  $\mathbf{0.934 \pm 0.003}$ \\
Pop  &  $0.957 \pm 0.002$  &  $0.845 \pm 0.004$  &  $0.866 \pm 0.004$  &  $0.899 \pm 0.005$  &  $0.883 \pm 0.004$  &  $\mathbf{0.939 \pm 0.004}$ \\
\bottomrule
\end{tabular}
}
\caption{Average Overlapping Area (OA) across seven music attributes for unconditional generation, with highest non-GT OA bolded.}
\label{tab:uncond_gen}
\end{table*}

\begin{table*}[h]
\centering
\resizebox{0.95\textwidth}{!}{
\begin{tabular}{lcccccc}
\toprule
Method & Loss $\downarrow$ (PH) & OA $\uparrow$ (PH) & Loss $\downarrow$ (ND) & OA $\uparrow$ (ND) & Loss $\downarrow$ (CP) & OA $\uparrow$ (CP) \\
\midrule
No Guidance & $0.018 \pm 0.010$ & $0.842 \pm 0.012$ & $2.486 \pm 3.530$ & $0.830 \pm 0.016$ & $0.831 \pm 0.142$ & $\mathbf{0.854 \pm 0.026}$ \\
Classifier & $0.005 \pm 0.004$ & $\underline{0.855 \pm 0.020}$ & $\underline{0.698 \pm 0.587}$ & $\mathbf{0.861 \pm 0.025}$ & $0.723 \pm 0.200$ & $0.850 \pm 0.033$ \\
DPS - NN & $\mathbf{0.001 \pm 0.002}$ & $0.849 \pm 0.018$ & $1.261 \pm 2.340$ & $0.667 \pm 0.113$ & $\underline{0.414 \pm 0.256}$ & $0.839 \pm 0.039$ \\
DPS - Rule & $0.010 \pm 0.008$ & $0.635 \pm 0.006$ & $2.508 \pm 2.798$ & $0.800 \pm 0.080$ & - & - \\
\ours (ours) & $\underline{0.003 \pm 0.004}$ & $\mathbf{0.867 \pm 0.005}$ & $\mathbf{0.131 \pm 0.325}$ & $\underline{0.842 \pm 0.031}$ & $\mathbf{0.273 \pm 0.1637}$ & $\underline{0.851 \pm 0.011}$ \\
\bottomrule
\end{tabular}
}
\caption{
Objective evaluation for individual rule guidance. Bottom 2 losses and top 2 OA metrics are highlighted. 
\ours significantly improves the controllability of non-differentiable rules.}
\label{tab:single_rule}
\end{table*} 

\subsection{Individual Rule Guidance}
\label{sec:single_rule}
\textbf{Setup.}
We consider three rules: pitch histogram, note density (vertical and horizontal) and chord progression, where pitch histogram is differentiable and the other two are non-differentiable (see Appendix \ref{sec:rule_def} for the full definition of each rule). 
In our evaluation of the guidance performance, we default to conditioning on the Muscore dataset unless otherwise specified, owing to its comprehensive variety and extensive coverage of a broad spectrum of rule labels.
For each rule, we randomly select 200 samples from the test dataset, and extract their attributes as the target for guided generation. We choose the number of samples to be 16 for \ours if without explicit mentioning.

\textbf{Baselines.}
We compare with two popular post-hoc guidance methods: classifier guidance \cite{dhariwal2021diffusion} and Diffusion Posterior Sampling (DPS) \cite{chung2022diffusion}. 
For classifier guidance, we train a classifier on noisy latent and target pair for each rule. DPS only requires the loss to be defined on clean data $\bx{0}$ so we can directly plug in the rule in the loss if the rule is differentiable (DPS-Rule) without any additional training. However, it still requires the gradient of the rule, and therefore we train a surrogate model (a neural network) for non-differentiable rules (DPS-NN). 

\textbf{Metrics.}
We evaluate conditional generation performance by two metrics: loss and OA. Loss reflects controllability: whether the generated samples follow the target rules. For pitch histogram and note density, we use L2 loss, and for chord progression, we use 0-1 loss. 
However, a low loss does not necessarily indicates good quality music. For instance, a music piece that follows the note density requirement may have random pitch and does not sound good. Therefore, we use the OA to measure how close the generated music distribution and the ground truth music distribution (with similar target attributes as the generated music) when projected to some music features space. Please see Appendix \ref{sec:appen_obj_eval_setup} for the detailed evaluation setup.

\textbf{Results.}
Table \ref{tab:single_rule} shows the results. 
We make three main observations. First, our method significantly outperforms the other methods in generating samples that follow the \emph{non-differentiable} rules (note density and chord progression). It achieves the lowest loss, without need for training any surrogate model, which is mandatory for classifier guidance and DPS-NN.
However, it sacrifices OA a bit. This is because over-optimizing for one attribute will overlook the other attributes. Reducing the number of samples used for SCG can lead to better OA at the cost of a higher loss (See Section \ref{sec:abla} on this trade-off).

Second, we find it challenging to train neural network surrogate models to approximate non-differentiable rules (Appendix \ref{sec:appen_cls_train}), leading to poor performance of guidance methods that rely on surrogate models. 
For differentiable rules (pitch histogram), the surrogate model learns well and DPS-NN achieves the lowest loss.

Third, to the best of our knowledge, our method is the first plug-and-play guidance method that supports non-differentiable and black-box loss functions. 
In contrast, DPS-rule fails to guide on note density because the gradient is zero almost everywhere. It also does not apply to chord progression because the loss involves a black-box API that cannot be back-propagated through.
Overall, our method proves especially beneficial for guiding the generation process with non-differentiable loss functions, or for achieving guidance without the need for additional training.

\subsection{Composite Rule Guidance}
\label{sec:multiple_rule}

We apply our method to generate samples that follow composite rules, following the same setup in Section \ref{sec:single_rule}, and assuming that the rule labels are conditionally independent given the sample. 
For classifier and DPS-NN, we train a surrogate model for each rule, and combine the gradient of each classifier to obtain the guidance term: $\Sigma_i \omega_i \nabla_{\bx{t}} \log p_t(\by_i | \bx{t})$.
For our method, we use a weighted loss function $\Sigma_i \omega_i \ell_{\by_i}(\bx{})$ to select the best direction for each step. 
We set the weights on pitch histogram, note density and chord progression to be 40, 1, 1 respectively so that their loss is on the same order of magnitude.  
In addition, we compared with four established conditional music generation baselines: Retrieval, Figaro-expert, Figaro-expert-learned \cite{von2022figaro} and MuseCOCO \cite{lu2023musecoco}. 
The Retrieval baseline is: given a set of target attributes, we find the sample within the datasets that has the closest attributes to the target attributes. This serves as an oracle baseline. 

\begin{table}[h]
\centering
\resizebox{\columnwidth}{!}{
\begin{tabular}{lcccc}
\toprule
Method & PH $\downarrow$ & ND $\downarrow$ & CP $\downarrow$ & OA $\uparrow$ \\
\midrule
Retrieval & $0.006 \pm 0.005$ & $0.433 \pm 1.068$ & $0.556 \pm 0.182$ & $\mathbf{0.886 \pm 0.005}$ \\
Figaro expert & $0.007 \pm 0.007$ & $2.303 \pm 2.256$ & $0.761 \pm 0.187$ & $0.857 \pm 0.047$ \\
Figaro expert+learned & $0.006 \pm 0.009$ & $1.489 \pm 2.737$ & $0.726 \pm 0.214$ & $0.883 \pm 0.008$ \\
MuseCoCo & $0.040 \pm 0.026$ & $2.734 \pm 3.551$ & $0.821 \pm 0.163$ & $0.753 \pm 0.038$ \\
\midrule
No Guidance & $0.018 \pm 0.010$ & $2.486 \pm 3.530$ & $0.831 \pm 0.142$ & $0.803 \pm 0.096$ \\
Classifier & $0.006 \pm 0.006$ & $0.822 \pm 0.844$ & $0.724 \pm 0.205$ & $0.859 \pm  0.026$ \\
DPS-NN & $0.004 \pm 0.006$ & $1.366 \pm 2.265$ & $0.661 \pm 0.257$ & $0.752 \pm 0.079$ \\
SCG & $0.014 \pm 0.009$ & $0.466 \pm 0.648$ & $0.446 \pm 0.205$ & $\mathbf{0.874 \pm 0.007}$ \\
DPS-NN + SCG & $\mathbf{0.002 \pm 0.007}$ & $0.238 \pm 0.531$ & $0.313 \pm 0.231$ & $0.781 \pm 0.084$ \\
Classifier + SCG & $0.003 \pm 0.005$ & $\mathbf{0.148 \pm 0.203}$ & $\mathbf{0.284 \pm 0.197}$ & $0.864 \pm 0.010$ \\
\bottomrule
\end{tabular}
}
\vspace{-0.1in}
\caption{Objective evaluation for composite rule guidance. \ours + Classifier achieves significantly lower losses for all three rules simultaneously with a high OA score.}
\label{tab:multi_rule}
\end{table}

The results are presented in Table \ref{tab:multi_rule}. 
We can see that, overall, our method achieves the lowest loss among all the baselines. The OA ranking is also higher than that of individual rule guidance, because guiding the generation with three attributes prevents over-optimizing for one attribute. There are two baselines that have higher OA than us: retrieval and Figaro-expert-learned. It is as expected that retrieval has the highest OA because the samples are selected from the dataset rather than being generated. Figaro-expert-learned also achieves high OA, because it is a style transfer model that needs to take in source music rather than a generative model. The generated music is conditioned on the latent representation of the source music. Therefore, it is not a fair comparison but we still put it here for reference. MuseCOCO cannot generate well following the target rules because it does not support fine-grained control.

Among the diffusion guidance methods, our method achieves much lower loss on non-differentiable rules compared to other methods, similar to the case of individual rule guidance. However, it compromises control over the pitch histogram. Additionally, the loss associated with each rule is higher than in scenarios of individual rule guidance, which aligns with expectations. This increase in loss occurs because it is more challenging to identify a direction that satisfies multiple rules simultaneously, as opposed to a single rule, within the same computational budget.

To enhance the controllability of composite rules, we integrate our \ours approach with gradient-based guidance methods. In this framework, the gradient of the surrogate model provides a preliminary guidance signal. \ours then identifies the optimal directions along these initially guided trajectories. As indicated in Table \ref{tab:multi_rule}, this combination of our method with the baseline gradient method results in improved controllability for each rule, compared to the baseline method alone. Furthermore, we achieve a level of controllability comparable to that of individual rule guidance, while using the same number of samples.

\subsection{Ablation Studies}
\label{sec:abla}
\textbf{Controllability and Computational Time Trade-off.}
Table \ref{tab:abla_bf} shows the loss achieved by \ours with different number of samples at each step. The time is reported for generating 4 samples in a batch.
As anticipated, more samples results in lower loss, but requires more time. 
To achieve a balance between controllability and computational efficiency, we integrate classifier guidance with \ours. This combination yields interesting results: number of samples of 4, when used in conjunction with classifier guidance, delivers similar performance to number of samples of 16 with \ours alone, but is approximately four times faster.

\textbf{Controllability and OA Trade-Off.} We observe a trade-off between controllability (measured by loss) and OA (used as a necessary, albeit not sufficient, indicator of music quality.) in Table \ref{tab:abla_bf}, where we guide the model to generate music following given note density. 
To better evaluate the trade-off, we compute 2 OA metrics, one uses the full dataset as reference (denoted by `OA full'), the other uses the data that comply with the desired rule (denoted by `OA' as in previous sections). The motivation of `OA full' stems from our approach of extracting note density from a diverse selection of sources within the dataset for conditional generation. If the generated pieces precisely mirror their source, they can achieve a low loss because they comply with the rule perfectly, and have a high OA because their sources are taken from the dataset (the `source’ row in Table \ref{tab:abla_bf}). 

We highlight the highest OA for each group of methods (the baselines, SCG and classifier+SCG with different number of samples). Note that the main difference between these two metrics is for `No Guidance', where `OA cluster' is much worse than `OA full'. This is because the `OA cluster' metric rewards controllability more. However the controllability and OA trade-off for SCG related methods are consistent among the two OA metrics.  
We think that this trade-off is caused by over-optimizing over some constraints. For instance, a generative model could generate music that follows the note density exactly, but completely ignore the pitch of the notes. One can balance controllability and OA by tuning the number of samples at each step. 

\begin{table}[h]
\centering
\resizebox{\columnwidth}{!}{
\begin{tabular}{lccccc}
\toprule
Method & $n$ & Loss $\downarrow$ & OA full $\uparrow$ & OA $\uparrow$ & Time (s) \\
\midrule
No Guidance & - & $2.486 \pm 3.530$ & $0.918 \pm 0.005$ & $0.830 \pm 0.016$ & 25.4 \\
Source & - & 0 & $\mathbf{0.923 \pm 0.008}$ & - & - \\
Classifier & 1 & $0.698 \pm 0.587$ & $0.914 \pm 0.006$ & $\mathbf{0.861 \pm 0.025}$ & 47.8 \\
DPS-NN & 1 & $1.261 \pm 2.340$ & $0.735 \pm 0.012$ & $0.667 \pm 0.113$ & 109.3 \\
\midrule
 & 4 & $0.318 \pm 0.770$ & $\mathbf{0.895 \pm 0.006}$ & $\mathbf{0.873 \pm 0.023}$ & 277.7 \\
\ours & 8 & $0.214 \pm 0.368$ & $0.877 \pm 0.006$ & $0.847 \pm 0.014$ & 531.6 \\
 & 16 & $\mathbf{0.131 \pm 0.325}$ & $0.880 \pm 0.003$ & $0.842 \pm 0.031$ & 1242.6 \\
\midrule
 & 4 & $0.151 \pm 0.298$  & $\mathbf{0.906 \pm 0.006}$ & $\mathbf{0.861 \pm 0.011}$ & 301.9 \\
Classifier + \ours & 8 & $0.098 \pm 0.179$ & $0.893 \pm 0.004$ & $0.839 \pm 0.024$ & 555.6 \\
 & 16 & $\mathbf{0.064 \pm 0.159}$ & $0.899 \pm 0.007$ & $0.849 \pm 0.018$ & 1253.9 \\
\bottomrule
\end{tabular}
}
\vspace{-0.1in}
\caption{Trade-offs between controllability, OA and computational time. $n$ refers to number of samples at each step. }
\label{tab:abla_bf}
\end{table}

\textbf{Impact of Sampling Strategy.}
By default, we use DDPM with 1000 steps as the base sampling algorithm, and apply \ours for rule guidance after 250 steps ($t=750$). The reason that we do not start \ours from the beginning is that the decoded piano rolls at the beginning are quite sparse after thresholding the background. Consequently, the losses between the generated attributes and target attributes are almost the same among different realizations at this stage, making it ineffective for selecting the best directions. 

To reduce the computational cost, we explore various sampling strategies, as detailed in Table \ref{tab:abla_sampling}. 
Firstly, we experimented with applying \ours intermittently, every $k$ steps ($k=2,5$), and specifically during either the initial phase (750-400) or the latter phase (400-0) of the DDPM-1000 process. Among these variants, conducting \ours every 2 steps yielded the lowest loss. While the loss remains higher than in our default setting, this approach is about twice as fast.
Additionally, we observed that applying \ours during the early phase of the process is more effective than in the later phase, likely due to greater perturbations early on, which enhance the likelihood of identifying optimal directions (see Appendix \ref{sec:appen_inter_loss} for more details).

Secondly, we considered early stopping of the DDPM-1000 process after $k$ steps ($k=800, 700$). This is motivated by our use of post-processing techniques like thresholding and smoothing note velocity on piano rolls, which reduces the need for fine-tuning in the latter stages of the generation process. Early stopping at 200 steps resulted in only marginally inferior outcomes but cut computational time by a quarter.

Finally, we explore the compatibility of \ours with other popular sampling algorithm for diffusion models, such as DDIM \cite{song2020denoising}. By default, DDIM is deterministic. However, our \ours algorithm needs stochasticity to search for the best direction. Therefore, we set stochasticity $\eta=1$ in the DDIM algorithm and refer the modified algorithm as stochastic DDIM (sDDIM). We tested sDDIM with 100, 50 and 25 steps. More steps offers lower loss and better music quality at a cost of longer sampling time.

\begin{table}[t]
\centering
\resizebox{\columnwidth}{!}{
\begin{tabular}{llccc}
\toprule
Method & Guided Steps & Loss $\downarrow$ & OA $\uparrow$ & Time (s) \\
\midrule
\multirow{6}{*}{DDPM-1000} & 750-0 & $\mathbf{0.131 \pm 0.325}$ & $0.880 \pm 0.003$ & 1242.6 \\
& every 2 & $0.365 \pm 0.559$ & $0.893 \pm 0.006$ & 635.4 \\
& every 5 & $0.632 \pm 0.577$ & $0.879 \pm 0.005$ & 269.8 \\
& 750-400 & $0.458 \pm 0.647$ & $0.902 \pm 0.009$ & 594.7 \\
& 400-0 & $1.297 \pm 1.772$ & $\mathbf{0.912 \pm 0.007}$ & 674.6 \\
\midrule
DDPM$^\dagger$-800 & 750-200 & $\mathbf{0.183 \pm 0.341}$ & $\mathbf{0.864 \pm 0.005}$ & 912.6 \\
DDPM$^\dagger$-700 & 750-300 & $1.950 \pm 1.344$ & $0.737 \pm 0.011$ & 747.3 \\
\midrule
sDDIM-100 & all & $\mathbf{0.303 \pm 0.509}$ & $\mathbf{0.887 \pm 0.005}$ & 164.3 \\
sDDIM-50 & all & $0.372 \pm 0.915$ & $0.879 \pm 0.008$ & 81.9 \\
sDDIM-25 & all & $0.428 \pm 0.683$ & $0.859 \pm 0.005$ & 40.7 \\
\bottomrule
\end{tabular}
}
\vspace{-0.1in}
\caption{Impact of sampling strategy. The numbers that follow the method names are the total sampling steps. $^\dagger$: early stopping. 
}
\label{tab:abla_sampling}
\end{table}

\subsection{Subjective Evaluation}
\label{sec:subj_test}
\begin{figure}
    \centering
    \includegraphics[width=0.975\linewidth]{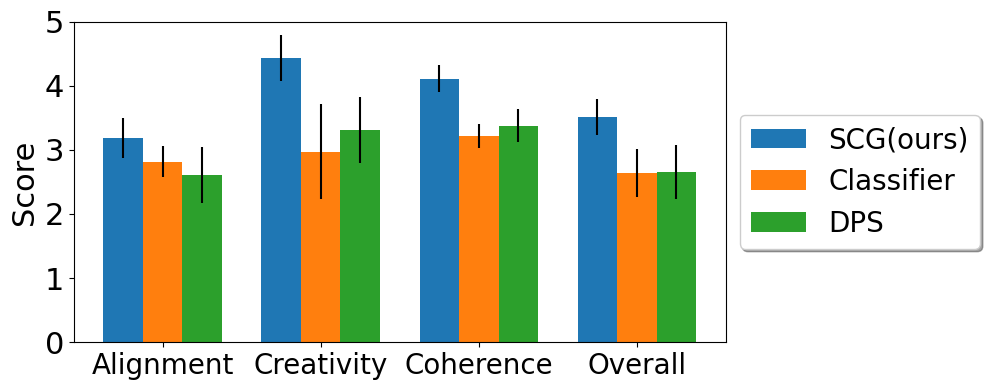}
    \vspace{-0.1in}
    \caption{Subjective evaluation scores. }
    \label{fig:survey}
    \vspace{-0.05in}
\end{figure}

To compare performance of our SCG algorithm and baselines (classifier guidance and DPS), we carried out a listening test. We crafted four sets of rules (each set comprised of PH, ND, and CP), and use each method to generate samples that follow the rules, yielding a total of 12 samples, each 10.24 seconds long. Experienced listeners assess the quality of samples in 4 dimensions: rule alignment, musical creativity, musical coherence, and overall rating. In Figure \ref{fig:survey}, SCG  consistently outperforms the baselines in all dimensions. For details of our survey, please see Appendix \ref{sec:rule-survey}.

\subsection{Examples of Our System as a Compositional Tool}
\label{sec:compositionExamples}
To demonstrate how our system can be used effectively as a compositional tool, we provide links to three example videos, available through \href{https://scg-rule-guided-music.github.io/}{this website}. For each video, a musician first indicated desired musical characteristics in terms of the rules (e.g. fairly sparse excerpt, following a simple I-V chord progression in C major, etc). The musician's plan was to then loop this and use that as an accompaniment over which they would then improvise. The system generated 3-5 options (i.e. samples) for each such request, and the musician chose their preferred sample with which to work, from each set. They sent the generated MIDI output to a Disklavier piano and then recorded a second track over top of that. In the accompanying videos, the musician is playing along with the generated music by our approach.

In \textbf{Video 1}, the model generated material that suggested a melody. Since the musician wanted to play the second track in the upper register, they first allowed the excerpt to play in full, as generated, and then removed the upper notes from the accompaniment to give room for themselves to play overtop. They chose to use the model's generated melody as a motif, and further improvise based on it.

In \textbf{Video 2}, the model generates an excerpt with a steady accompanying triplet ostinato behind a slower-moving descending melody in C minor (that suggests a progression that moves between the I and the V).

In \textbf{Video 3}, the model generates a sample with a changing note density and texture, and a slightly ambiguous harmonic quality that allowed flexibility in the improvising over it.
\section{Conclusion and Future directions}
We introduced a symbolic music generator with non-differentiable rule guided diffusion models,  drawing inspiration from stochastic control. Comprehensive evaluations show our model's superiority over previous works, highlighting the potential of rule-guided approaches in enhancing creativity and control in computational music composition.

In principle, the SCG algorithm introduced in this paper extends beyond the realm of symbolic music generation. Its capability to enforce diffusion models to follow non-differentiable constraints makes it suitable for diverse fields, as long as the constraints can be programmed and one can define a loss on how well the constraints are satisfied. For instances, in protein design, one can write a function to return how specific topological constraints are satisfied. In astronomy imaging, one can use black-box physics simulators as the rule function. We believe it is an exciting future direction to extend this algorithm to a broader scope.

\section*{Acknowledgements}
This project is funded in part by AeroVironment and NSF \#1918655.  Ziniu Hu is also funded in part by a Caltech CDSF Postdoctoral Fellowship. Sageev Oore and Chandramouli Sastry also thank the Canadian Institute for Advanced Research (CIFAR) and NSERC for their support.

\section*{Impact Statement}
This paper presents work whose goal is to advance the field of generative modeling for symbolic music. While there are numerous potential societal implications associated with our work, we believe none require specific emphasis in this context.

\bibliography{refs}
\bibliographystyle{icml2024}

\newpage
\appendix
\onecolumn
\section{Proofs.}
\label{sec:proof}
\subsection{Proof of Theorem \ref{thm:path_int_guidance}.}
\label{sec:thm1_proof}
\begin{lemma}[\citealp{dai1991stochastic,pavon1989stochastic}]
\label{lemma}
The transition probability for the stochastic dynamical system Eq \ref{eq:controlled_sde} with cost Eq \ref{eq:cost} and optimal control $\bu{}^*$ is:
\begin{equation}
    \mQ^*_{s,t}(\scstate{s}, \scstate{t}) = \mQ^0_{s,t}(\scstate{s}, \scstate{t})\frac{\Psi(\scstate{t},t)}{\Psi(\scstate{s},s)}
\end{equation}
where $\mQ^*_{s,t}(\scstate{s}, \scstate{t})$ denotes the transition probability from state $\scstate{s}$ at time $s$ to state $\scstate{t}$ at time $t$, and $\mQ^0_{s,t}(\scstate{s}, \scstate{t})$ denotes the transition probability of the uncontrolled system Eq \ref{eq:uncontrolled_sde}.
\end{lemma}

Now we prove Theorem \ref{thm:path_int_guidance}.
\begin{proof}
    Consider the SDE in Eq \ref{eq:controlled_sde} with initial condition $\scstate{0} \sim \delta_{\initial}$, where $\delta_{\initial}$ is a Dirac distribution centered at $\initial$. Define the terminal cost to be $\terminal{\scstate{T}} = \log \frac{\diracp(\scstate{T})}{\diracp(\scstate{T} | \by)}$, where $\diracp(\scstate{T})$ denotes the terminal distribution of the uncontrolled SDE in Eq \ref{eq:uncontrolled_sde} with initial condition $\scstate{0} \sim \delta_{\initial}$, and the target terminal distribution $\diracp(\scstate{T} | \by) := p(\by | \scstate{T}) \diracp(\scstate{T}) / p(\by)$.
    Then we have
    \begin{align}
        & \Psi(\initial, 0) = \mathbb{E}_{\mQ^0} \left[ e^{-\terminal{\scstate{T}}} | \scstate{t}=\initial \right] = \int_{\scstate{T}} \frac{\diracp(\scstate{T} | \by)}{\diracp(\scstate{T})} \diracp(\scstate{T}) \dd \scstate{T} = 1 \\
        & \Psi(\scstate{T}, T) = e^{-\terminal{\scstate{T}}} = \frac{\diracp(\scstate{T} | \by)}{\diracp(\scstate{T})}
    \end{align}
    Then from Lemma \ref{lemma}, we have 
    \begin{align}
        \mQ^*_{0, T}(\initial, \scstate{T}) &= \mQ^0_{0, T}(\initial, \scstate{T}) \frac{\Psi(\scstate{T}, T)}{\Psi(\initial, 0)} \notag \\
        &= \diracp(\scstate{T}) \frac{\diracp(\scstate{T} | \by)}{\diracp(\scstate{T})} \notag \\
        &= \diracp(\scstate{T} | \by)
    \end{align}
    From the properties of reverse-time SDE, we know that if $p_0(\scstate{}) = \mathcal{N}(\mathbf{0}, \mathbf{I})$, then $p_T(\scstate{})=p(\scstate{})$, i.e. $\int \diracp(\scstate{T}) p_0(\initial) \dd \initial = p(\scstate{T})$. It follows that
    \begin{align}
        \mQ^*(\scstate{T}) &= \int \mQ^*_{0, T}(\initial, \scstate{T}) p_0(\initial) \dd \initial \notag \\
        &= \int \diracp(\scstate{T} | \by) p_0(\initial) \dd \initial \notag \\
        &= \int \frac{p(\by | \scstate{T}) \diracp(\scstate{T})}{p(\by)} p_0(\initial) \dd \initial \notag \\
        &= \frac{p(\by | \scstate{T})}{p(\by)} \int \diracp(\scstate{T}) p_0(\initial) \dd \initial \notag \\
        &= \frac{p(\by | \scstate{T})}{p(\by)} p(\scstate{T}) \notag \\
        &= p(\scstate{T} | \by)
    \end{align}
    Finally, we show that the optimal control for $\tilde{\terminalonly} = \ell_y(\scstate{T})$ is the same as that for $\terminal{\scstate{T}} = \log \frac{\diracp(\scstate{T})}{\diracp(\scstate{T} | \by)}$, because
    \begin{equation}
    \label{eq:terminal_cost}
        \terminal{\scstate{T}} = \log \frac{\diracp(\scstate{T})}{\diracp(\scstate{T} | \by)} = \log \frac{p(\by)}{p(\by | \scstate{T})} = - \log p(\by | \scstate{T}) + const = \ell_y(\scstate{T}) + const 
    \end{equation}
    The last equality follows from our assumption that $p_0(\by|\scstate{}) \propto e^{-\ell_{\by}(\scstate{})}$. 
    Plugging $\tilde{\terminalonly}(\scstate{T})$ and $\terminal{\scstate{T}}$ into Eq \ref{eq:optimal_u_pi} leads to the same optimal control.    
\end{proof}

\subsection{Proof of Proposition \ref{thm:desirability_func}}
\label{sec:thm2_proof}
\begin{proof}
    \begin{align}
    \label{eq:V_classifier}
        \Psi(\scstate{}, t) &= \mathbb{E}_{\mQ^0} \left[ e^{-\terminal{\scstate{T}}} | \scstate{t}=\scstate{} \right] \notag \\
        &= \mathbb{E}_{\mQ^0} \left[ p(\by | \scstate{T}) \cdot c | \scstate{t}=\scstate{} \right] \notag \\
        &= c \cdot \int p(\by | \scstate{T}) p(\scstate{T} | \scstate{t} = \scstate{}) \dd \scstate{T} 
    \end{align}
    where $p(\scstate{T} | \scstate{t} = \scstate{}) := \mQ^0_{t, T}(\scstate{}, \scstate{T})$ is the transition probability from state $\scstate{}$ at time $t$ to state $\scstate{T}$ at time $T$ for the uncontrolled SDE, and it is differentiable with respect to $\scstate{}$ for all $0 \leq t < T$.
    In addition, notice that in diffusion models,
    \begin{align}
    p(\by | \scstate{t}) &= \int p(\by | \scstate{T}, \scstate{t}) p(\scstate{T} | \scstate{t}) \dd \scstate{T} \notag \\
    &= \int p(\by | \scstate{T}) p(\scstate{T} | \scstate{t}) \dd \scstate{T}
    \end{align}
    where the second equality comes from that $\scstate{t}$ and $\by$ are conditionally independent given $\scstate{T}$. 
    Then from Eq \ref{eq:V_classifier}, we have $\Psi(\scstate{t}, t) = c \cdot p(\by | \scstate{t})$.
\end{proof}

\begin{remark}
    Although $\terminal{\scstate{T}}$ could be non-differentiable when we choose non-differentiable loss functions, the desirability function $\Psi(\scstate{}, t)$ is differentiable with respect to $\scstate{}$ for all $0 \leq t < T$.
\end{remark}

\section{Compatibility of \ours with Various Sampling Procedures}
\label{sec:appen_algo}
\ours is compatible with many stochastic sampling procedures in diffusion models. The key of \ours is to sample multiple $\bx{t-1}$ given $\bx{t}$, and select the one that leads to the sample that follows the rule best. One can choose different sampling procedure to obtain $\bx{t-1}$ given $\bx{t}$.
Algorithm \ref{alg:editing_algo} shows how to use \ours with replacement-based editing. 
Algorithm \ref{alg:sampling_algo_ddim} shows how to use \ours with stochastic DDIM \cite{song2020denoising}.

\begin{algorithm}
   \caption{Editing with \ours.}
   \label{alg:editing_algo}
\begin{algorithmic}
    \STATE \textbf{Require:} Encoding of the source music $\tilde{\bx{}}_0$, mask  $\mask$ (1 for unaltered part and 0 for editing region), noise level $K$, sampling algorithm (e.g. $\operatorname{\ours}$), desired label $\by$ (optional, do not need if want to create a variant).
   \STATE $\mathbf{z} \sim \mathcal{N}(\mathbf{0}, \mathbf{I})$
   \STATE $\bx{K} = \sqrt{\bar{\alpha}_K} \tilde{\bx{}}_0 + \sqrt{1-\bar{\alpha}_K} \mathbf{z}$
   \FOR{$t=K$ {\bfseries to} $1$}
        \STATE $\triangleright$ Estimate the clean sample from noisy sample.
        \STATE $\hat{\mathbf{x}}_0 = \frac{1}{\sqrt{\bar{\alpha}_{t}}}\left(\mathbf{x}_{t} - \sqrt{1-\bar{\alpha}_{t}}\epsilon_\theta\left(\mathbf{x}_{t}, t\right)\right)$
        \STATE $\triangleright$ Replacement projection based on the mask.
        \STATE $\tilde{\bx{}}_0$ = $\mask \odot \bx{0} + (\mathbf{I} -\mask) \odot \hat{\mathbf{x}}_0$
        \STATE $\triangleright$ Predict $\mathbf{\epsilon}$ from $\tilde{\bx{}}_0$.
        \STATE $\tilde{\mathbf{\epsilon}} = \frac{1}{\sqrt{1 - \bar{\alpha}_t}}(\bx{t} - \sqrt{\bar{\alpha}_t} \tilde{\bx{}}_0 )$
        \STATE $\triangleright$ Sampling using corrected $\mathbf{\epsilon}$.
        \STATE $\bx{t-1} = \operatorname{sampling\_algorithm}(\bx{t}, t, \mathbf{\epsilon}, \by)$
   \ENDFOR
   \STATE {\bfseries return: } $\bx{0}$
\end{algorithmic}
\end{algorithm}

\begin{algorithm}
   \caption{Stochastic Control Guided stochastic DDIM sampling}
   \label{alg:sampling_algo_ddim}
\begin{algorithmic}
   \STATE \textbf{Require:} Loss function $\ell_y$, rule target $\by$, number of samples $n$, stochasticity $\eta > 0$, number of steps $S$. 
   \STATE $\bx{T} \sim \mathcal{N}(\mathbf{0}, \mathbf{I})$
   \FOR{$s=S$ {\bfseries to} $1$}
        \STATE $\triangleright$ Compute the posterior mean of $\bx{\tau_{s-1}}$.
        \STATE $\sigma_{\tau_s} = \eta \sqrt{\frac{1-\bar{\alpha}_{\tau_{s-1}}}{1-\bar{\alpha}_{\tau_s}}(1-\frac{\bar{\alpha}_{\tau_s}}{\bar{\alpha}_{\tau_{s-1}}})}$
        \STATE $\hat{\bx{}}_{\tau_{s-1}} = \sqrt{\bar{\alpha}_{\tau_{s-1}}}\left( \frac{\bx{\tau_s} - \sqrt{1-\bar{\alpha}_{\tau_s}}\epsilon_\theta\left(\mathbf{x}_{\tau_s}, \tau_s\right)}{\sqrt{\bar{\alpha}_{\tau_s}}} \right) + \sqrt{1 - \bar{\alpha}_{\tau_{s-1}}-\sigma_{\tau_s}^2}\epsilon_\theta\left(\mathbf{x}_{\tau_s}, \tau_s\right)$
        \IF{$s>1$}
            \STATE $\triangleright$ Sampling possible next steps.
            \STATE $\mathbf{x}_{\tau_{s-1}}^i = \hat{\bx{}}_{\tau_{s-1}} + \sigma_{\tau_s} \mathbf{z}^i$, with $\mathbf{z}^1, ..., \mathbf{z}^n \sim \mathcal{N}(\mathbf{0}, \mathbf{I})$
            \STATE $\triangleright$ Estimate the clean sample from noisy sample.
            \STATE $\hat{\mathbf{x}}_0^i = \frac{1}{\sqrt{\bar{\alpha}_{\tau_{s-1}}}}\left(\mathbf{x}_{\tau_{s-1}}^i - \sqrt{1-\bar{\alpha}_{\tau_{s-1}}}\epsilon_\theta\left(\mathbf{x}^i_{\tau_{s-1}}, \tau_{s-1}\right)\right)$
            \STATE $\triangleright$ Find the direction that minimizes the loss.
            \STATE $k = \argmax_i \log p(\by | \hat{\bx{}}_0^i ) = \argmax_i -\ell_y(\hat{\bx{}}_0^i)$
            \STATE $\mathbf{x}_{\tau_{s-1}} = \mathbf{x}_{\tau_{s-1}}^k$
        \ELSE
            \STATE $\mathbf{x}_{\tau_{s-1}} = \hat{\bx{}}_{\tau_{s-1}}$
        \ENDIF
   \ENDFOR
   \STATE {\bfseries return: } $\bx{0}$
\end{algorithmic}
\end{algorithm}

\section{Additional Experiment Results}
\label{sec:appen_additional_exp}
\subsection{Unconditional Generation}
In Table \ref{tab:detail_uncond_gen}, we report the overlapping area between the intra-set and inter-set distribution for all 7 musical attributes as proposed in \cite{yang2020evaluation}. The highest and second highest value except for GT are high- lighted in bold and underline respectively. Our method achieves the highest average OA on all the datasets, and achieves the top 2 OA for most of the individual attributes.

\begin{table*}[h]
\centering
\resizebox{\textwidth}{!}{
\begin{tabular}{llcccccccc}
\toprule
Dataset &   Method &                    Used Pitch &                           IOI &                    Pitch Hist &                   Pitch Range &                      Velocity &                 Note Duration &                  Note Density &                           Avg \\
\midrule
\multirow{6}{*}{Maestro} &       GT &             $0.960 \pm 0.007$ &             $0.901 \pm 0.007$ &             $0.980 \pm 0.003$ &             $0.962 \pm 0.004$ &             $0.971 \pm 0.004$ &             $0.884 \pm 0.011$ &             $0.953 \pm 0.005$ &             $0.944 \pm 0.002$ \\
 &  MusicTr & $\underline{0.954 \pm 0.004}$ & $\underline{0.896 \pm 0.018}$ & $\underline{0.948 \pm 0.011}$ &             $0.961 \pm 0.005$ & $\underline{0.967 \pm 0.005}$ &             $0.647 \pm 0.022$ & $\underline{0.948 \pm 0.007}$ & $\underline{0.903 \pm 0.005}$ \\
 &     Remi &             $0.934 \pm 0.018$ &             $0.794 \pm 0.017$ &             $0.897 \pm 0.010$ &             $0.903 \pm 0.015$ &             $0.906 \pm 0.007$ &             $0.542 \pm 0.021$ &    $\mathbf{0.952 \pm 0.004}$ &             $0.847 \pm 0.005$ \\
 &      CPW &             $0.941 \pm 0.012$ &             $0.641 \pm 0.025$ &             $0.866 \pm 0.010$ & $\underline{0.962 \pm 0.005}$ &             $0.830 \pm 0.014$ &             $0.436 \pm 0.030$ &             $0.933 \pm 0.007$ &             $0.801 \pm 0.006$ \\
 & PolyDiff &             $0.852 \pm 0.006$ &             $0.843 \pm 0.026$ &             $0.888 \pm 0.016$ &             $0.871 \pm 0.006$ &             $0.805 \pm 0.031$ & $\underline{0.777 \pm 0.016}$ &             $0.856 \pm 0.005$ &             $0.842 \pm 0.007$ \\
 &     Ours &    $\mathbf{0.961 \pm 0.006}$ &    $\mathbf{0.901 \pm 0.009}$ &    $\mathbf{0.960 \pm 0.010}$ &    $\mathbf{0.963 \pm 0.006}$ &    $\mathbf{0.971 \pm 0.004}$ &    $\mathbf{0.910 \pm 0.012}$ &             $0.934 \pm 0.005$ &    $\mathbf{0.943 \pm 0.003}$ \\
 \midrule
\multirow{6}{*}{Muscore} &       GT &             $0.959 \pm 0.009$ &             $0.928 \pm 0.019$ &             $0.980 \pm 0.005$ &             $0.965 \pm 0.006$ &             $0.896 \pm 0.008$ &             $0.925 \pm 0.008$ &             $0.963 \pm 0.004$ &             $0.945 \pm 0.004$ \\
 &  MusicTr & $\underline{0.952 \pm 0.012}$ & $\underline{0.916 \pm 0.008}$ &             $0.891 \pm 0.009$ & $\underline{0.958 \pm 0.005}$ & $\underline{0.790 \pm 0.008}$ & $\underline{0.851 \pm 0.020}$ &    $\mathbf{0.949 \pm 0.005}$ & $\underline{0.901 \pm 0.004}$ \\
 &     Remi &             $0.924 \pm 0.008$ &    $\mathbf{0.917 \pm 0.016}$ & $\underline{0.955 \pm 0.010}$ &             $0.944 \pm 0.007$ &             $0.731 \pm 0.024$ &             $0.748 \pm 0.028$ &             $0.934 \pm 0.003$ &             $0.879 \pm 0.006$ \\
 &      CPW &             $0.879 \pm 0.012$ &             $0.839 \pm 0.020$ &             $0.941 \pm 0.008$ &             $0.900 \pm 0.006$ &             $0.750 \pm 0.025$ &             $0.715 \pm 0.030$ &             $0.879 \pm 0.010$ &             $0.843 \pm 0.007$ \\
 & PolyDiff &             $0.888 \pm 0.009$ &             $0.831 \pm 0.006$ &             $0.927 \pm 0.012$ &             $0.864 \pm 0.007$ &             $0.693 \pm 0.014$ &             $0.822 \pm 0.015$ &             $0.891 \pm 0.008$ &             $0.845 \pm 0.004$ \\
 &     Ours &    $\mathbf{0.963 \pm 0.004}$ &             $0.915 \pm 0.009$ &    $\mathbf{0.962 \pm 0.009}$ &    $\mathbf{0.964 \pm 0.004}$ &    $\mathbf{0.890 \pm 0.006}$ &    $\mathbf{0.900 \pm 0.012}$ & $\underline{0.946 \pm 0.005}$ &    $\mathbf{0.934 \pm 0.003}$ \\
 \midrule
\multirow{6}{*}{Pop} &       GT &             $0.956 \pm 0.007$ &             $0.949 \pm 0.006$ &             $0.983 \pm 0.002$ &             $0.955 \pm 0.003$ &             $0.954 \pm 0.004$ &             $0.940 \pm 0.009$ &             $0.963 \pm 0.002$ &             $0.957 \pm 0.002$ \\
     &  MusicTr &             $0.807 \pm 0.016$ &             $0.880 \pm 0.010$ &             $0.852 \pm 0.005$ &             $0.833 \pm 0.011$ & $\underline{0.865 \pm 0.014}$ &             $0.871 \pm 0.008$ &             $0.809 \pm 0.011$ &             $0.845 \pm 0.004$ \\
     &     Remi &             $0.870 \pm 0.014$ &             $0.839 \pm 0.007$ &    $\mathbf{0.979 \pm 0.002}$ &             $0.827 \pm 0.008$ &             $0.853 \pm 0.013$ &             $0.826 \pm 0.012$ &             $0.867 \pm 0.005$ &             $0.866 \pm 0.004$ \\
     &      CPW &             $0.921 \pm 0.011$ &             $0.803 \pm 0.022$ &             $0.942 \pm 0.010$ &             $0.927 \pm 0.008$ &             $0.853 \pm 0.006$ &             $0.891 \pm 0.011$ &    $\mathbf{0.953 \pm 0.008}$ & $\underline{0.899 \pm 0.005}$ \\
     & PolyDiff &    $\mathbf{0.941 \pm 0.003}$ & $\underline{0.924 \pm 0.012}$ &             $0.964 \pm 0.005$ &    $\mathbf{0.937 \pm 0.006}$ &             $0.648 \pm 0.007$ &    $\mathbf{0.912 \pm 0.020}$ &             $0.855 \pm 0.012$ &             $0.883 \pm 0.004$ \\
     &     Ours & $\underline{0.927 \pm 0.009}$ &    $\mathbf{0.952 \pm 0.004}$ & $\underline{0.969 \pm 0.002}$ & $\underline{0.928 \pm 0.013}$ &    $\mathbf{0.948 \pm 0.003}$ & $\underline{0.911 \pm 0.019}$ & $\underline{0.941 \pm 0.009}$ &    $\mathbf{0.939 \pm 0.004}$ \\
\bottomrule
\end{tabular}
}
\caption{Objecetive evaluation of unconditional generation. The overlapping area (OA) for 7 music attributes and the average OA are reported. The highest and second highest OA excluding GT are bolded and underlined respectively.}
\label{tab:detail_uncond_gen}
\end{table*}

In addition, we check if the models are copying from the dataset. To do so, we randomly pick 50 generated samples for each method and 2000 samples from the training dataset, and identify the closest musical piece to a generated MIDI file from the dataset. Specifically, we extract and compare key features from each file, including pitch, velocity, duration of notes, and overall note density. We report the average matching score of these four features in Table \ref{tab:matching_score}. As we can see, despite our method achieving the highest OA, it has the second lowest matching score, indicating that the OA improvement is not from copying from the dataset. 

\begin{table}[h]
\centering
\resizebox{0.3\textwidth}{!}{
\begin{tabular}{lcc}
\toprule
Method & Matching Score & OA \\
\midrule
MusicTr & $0.6273 \pm 0.0336$ & $0.903 \pm 0.005$ \\
Remi & $0.6386 \pm 0.0210$ & $0.866 \pm 0.004$ \\
CPW & $0.6439 \pm 0.0134$ & $0.899 \pm 0.005$ \\
PolyDiff & $0.6372 \pm 0.0097$ & $0.883 \pm 0.004$ \\
Ours & $0.6337 \pm 0.0185$ & $0.943 \pm 0.003$ \\
\bottomrule
\end{tabular}
}
\caption{Test if the model is copying samples from the dataset by evaluating Matching Score and OA.}
\label{tab:matching_score}
\end{table}

\subsection{Editing}
\label{sec:edit}
Our framework also supports editing. Given an existing music piece, we can modify it within any given time window: either create a new variant or guide it to satisfy new rules. 
To achieve this, we mainly follow the SDEdit framework \cite{meng2021sdedit}: first we add Gaussian noise of a chosen noise level to the latent music representation and then progressively remove the noise by reversing the SDE. During the reverse process, we use a mask to distinguish the parts that we want to preserve unaltered and the portion we want to modify, and we condition on the unaltered parts via replacement-based conditioning methods as in \cite{choi2021ilvr,kawar2022denoising}. 
Please refer to Appendix \ref{sec:appen_algo} for the detailed guided editing algorithm.

We benchmark our music editing performance against two estabilished methods: MuseMorphose \cite{wu2023musemorphose} and PolyDiffusion \cite{min2023polyffusion}. Since these baselines are trained on Pop piano music, we condition on the Pop dataset when evaluating our method. Unlike these baselines, which restrict editing to one specific attribute, our method offers the flexibility to edit any attribute. The editing task involves creating a new music piece that adheres to predefined rules based on an original source music piece (for detailed settings, see Appendix \ref{sec:appen_obj_eval_setup}). To assess controllability, we evaluate the error rate between the target and generated attributes. Additionally, we measure the similarity in chroma and groove between the generated piece and the source to gauge their resemblance. The goal is to generate music that not only complies with the desired rules but also closely resembles the original source music.

Table \ref{tab:edit} shows the results. For note density, we experiment with two noise levels: 400 and 500. For chord progression, we use a noise level of 500. We can see that there is a trade-off between controllability and resemblance: higher noise level results in better controllability (lower error) but reduced resemblance (lower similarity metrics).

\begin{table}[h]
\centering
\resizebox{0.6\columnwidth}{!}{
\begin{tabular}{llccc}
\toprule
Rule & Method & Error (\%) $\downarrow$ & $Sim_{chr} \uparrow$ & $Sim_{grv} \uparrow$  \\
\midrule
\multirow{3}{*}{Note Density} & MuseMorphose & 29.34 & $\mathbf{0.9130}$ & $\mathbf{0.9184}$ \\
& Ours-400 & 35.62 & 0.9119 & 0.8511 \\
& Ours-500 & $\mathbf{27.87}$ & 0.8173 & 0.7153 \\
\midrule
\multirow{2}{*}{Chord Progression} & PolyDiffusion & 70.48 & 0.5902 & $\mathbf{0.7515}$ \\
& Ours & $\mathbf{12.62}$ & $\mathbf{0.8236}$ & 0.6974 \\
\bottomrule
\end{tabular}
}
\caption{Editing performance. For note density, we experimented with noise level of 400 and 500. For chord progression, we used noise level of 500.}
\label{tab:edit}
\end{table}

\section{Detailed Experiment Setup}
\subsection{Music Rules}
\label{sec:rule_def}
We consider three music rules and give their definitions below.

\textbf{Pitch Histogram}: We compute the histogram of 12 pitch classes over the whole piano roll. Pitch velocity is considered when computing the histogram. The target $\by$ is a 12-dimensional vector specifying the desired pitch histogram.

\textbf{Note Density}: We control both vertical and horizontal note density. We compute note density within $128 \times 128$ windows. For a piano roll of shape $128 \times 1024$, the target $\by$ is a 16-dimensional vector, the first 8 dimension are for vertical note density and the last 8 dimension are for horizontal note density.

Vertical note density $\text{ND}_{\text{vertical}}$ is computed by 
\begin{equation}
    \text{ND}_{\text{vertical}} = \frac{1}{T}\sum_{t=1}^T n_{\text{on}}(t)
\end{equation}
where $n_{\text{on}}(t)$ stands for the number of on-notes at time $t$, and $T$ is the window size, we set $T=128$.

Horizontal note density $\text{ND}_{\text{horizontal}}$ is computed by 
\begin{equation}
    \text{ND}_{\text{horizontal}} = \sum_{t=1}^T \mathbbm{1}(n_{\text{start}}(t) \geq 1)
\end{equation}
where $n_{\text{start}}(t)$ stands for the number of notes that start at time $t$.

\textbf{Chord Progression}: We extract chords using chord analysis tool from the $\operatorname{music21}$ \cite{cuthbert2010music21} package, and group them into 7 classes. We extract 8 chords in total for the $128 \times 1024$ piano roll, each chord is the longest chord within a $128 \times 128$ window. The target $\by$ is an 8-dimensional vector specifying the desired chord for each $128 \times 128$ window.

\subsection{Training Setup}
\label{sec:appen_training}
\textbf{Data Augmentation.} 
We use the same data augmentation for both VAE and diffusion model training.
\begin{itemize}
    \item \textbf{Key shift}: Entire piano rolls were shifted by up to 6 pitches, effectively functioning as a key switch.
    \item \textbf{Time shift}: We load in a piano roll of 2 times the desired length, and randomly select a starting time to obtain the actual piano roll for training. 
    \item \textbf{Tempo shift}: Tempo of the piece was shifted by a factor of [0.95, 1.05]. 
\end{itemize}

\textbf{VAE.}
Utilizing the standard autoencoder architecture from \cite{rombach2022high}, we compressed piano roll segments (dimension $3 \times 128 \times 128$) into a latent space of $4 \times 16 \times 16$. The three dimensions of the piano roll include onset and pedal information, in addition to the standard piano roll data.

Let $x$ represent the piano roll in pixel space and $z$ the latent code. We denote the encoder and decoder as $\mathcal{E}$ and $\mathcal{D}$, respectively. The training objective for our VAE model is formulated as follows:
\begin{equation}
    \mathcal{L}_{VAE} = \| \mathcal{D}(\mathcal{E}(x)) - x \|_1 + \lambda_{\text{KL}}(t) D_{KL}(\mathcal{N}(z; \mathcal{E}_{\mu}, \mathcal{E}_{\sigma^2}) \| \mathcal{N}(z; 0, I)) 
    + \lambda_{\text{denoise}}(t) \| \mathcal{D}(\mathcal{E}(\operatorname{Noisy}(x))) - x \|_1
\end{equation}

The first term is the standard reconstruction loss, where we used L1 loss to encourage sparsity. The second term is the standard KL regularization term weighted by a scheduler $\lambda_{\text{KL}}(t)$. 
The third term is a denoising reconstruction loss, influenced by the scheduler $\lambda_{\text{denoise}}(t)$. Here, $\operatorname{Noisy}$ refers to a perturbation operator applied to the piano roll, encompassing:
\begin{itemize}
    \item \textbf{Note shift}: Some fraction of notes were randomly selected by uniform distribution to be perturbed. Perturbed notes were shifted by up to 2 pitches higher or lower.
    \item \textbf{Adjacent note addition}: Some fraction of notes were randomly selected by uniform distribution. A second identical note was added just one pitch higher or lower to the original note. These adjacent notes are quite discordant to the ear.
    \item \textbf{Rhythm shift}: Some fraction of notes were randomly selected by uniform distribution to be perturbed. Perturbed notes were shifted by up to 100 ms earlier or later. 
    \item \textbf{Note deletion}: Some fraction of notes were randomly selected by uniform distribution to be deleted.
\end{itemize}
We capped the maximum fraction of perturbed notes at $25\%$ for all perturbations. The model was trained over 800k steps. The KL scheduler, $\lambda_{\text{KL}}(t)$, was a linear scheduler increasing from 0 to $1e-2$ across 400k steps. The denoising scheduler, $\lambda_{\text{denoise}}$, linearly increased the perturbation fraction from 0 to $25\%$ over 400k steps. We employed a cosine learning rate scheduler with a 10k-step warmup, peaking at a learning rate of $5e-4$. The optimizer used was AdamW \cite{loshchilov2017decoupled}, with weight decay of 0.01 and a batch size of 80.

\textbf{Diffusion Model.}
We train our diffusion model with a transformer backbone on the latent space of a pretrained VAE. First we rescale the latent representation $\mathcal{E}(x)$ by its standard deviation, computed using a batch of 256 training samples as per the methodology described in \cite{rombach2022high}. Then we reshaped the latent representation from $4 \times 16 \times 128$ to $32 \times 256$, followed by a transformation of the 32-dimensional vector to match the hidden dimension of the transformer backbone. We employed the DiT-XL architecture from \cite{peebles2023scalable}, which has a hidden dimension of 1152. In addition, we use rotary positional embedding \cite{su2023roformer} to accommodate for various length of input (e.g. when generating longer sequence of music).

Given our use of data augmentation during diffusion model training, it was necessary to compute $\mathcal{E}(x)$ dynamically, a process that is typically time-consuming. To optimize this, we employed a strategy to avoid encoding each sample from scratch. During data loading, we initially loaded a piano roll of length 2560 and then encoded each 128-length segment using the pretrained encoder, resulting in 20 latent codes. By concatenating subsets of these latent codes, we generated 4 training samples (segments 1-8, 5-12, 9-16, and 13-20), each measuring $8 \times 128 = 1024$ in length.

We train our model on three datasets using the training procedure of classifier-free guidance \cite{ho2022classifier}. Specifically, we set $y=0$ for Maestro, $y=1$ for Muscore and $y=2$ for Pop. We jointly train a conditional model $\epsilon_\theta\left(\bx{t}, t, y\right)$ and an unconditional model $\epsilon_\theta\left(\bx{t}, t, \text{null}\right)$ with a dropout rate of 0.1.

We adhered to the training hyper-parameters outlined in \cite{peebles2023scalable}. Specifically, we used a constant learning rate of $1e-4$, no weight-decay and a batch size of 256 with the AdamW optimizer \cite{loshchilov2017decoupled}. We use linear noise scheduling and trained the model for 1.2M steps.

\subsection{Objective Evaluation Setup}
\label{sec:appen_obj_eval_setup}
\textbf{Unconditional Generation.} 
In our study, we generated 400 music segments, each lasting 10.24 seconds, for all the methods under consideration. For baseline methods that utilize bars as the time unit, we produced 8-bar sequences from which we randomly extracted segments of 10.24 seconds in duration.
We used the official released models for Remi \cite{huang2020pop}, CPW \cite{hsiao2021compound} and PolyDiff \cite{min2023polyffusion}. 
Unfortunately, an official implementation for the music transformer \cite{huang2018music} was not available. Consequently, we resorted to an unofficial implementation\footnote{Available at \url{https://github.com/gwinndr/MusicTransformer-Pytorch}} and trained a music transformer by ourselves. 

The overlapping area (OA) for seven music attributes was computed following the methodology described in \cite{yang2020evaluation}. To accurately evaluate OA, it is typically required that the two datasets being compared contain an equal amount of data. Therefore, we randomly selected 400 samples from the test dataset to align with the number of generated samples. This evaluation process was repeated five times to calculate the mean and standard deviation of the results in Table \ref{tab:uncond_gen}.

\textbf{Individual Rule Guidance.}
For each rule, we randomly selected 200 samples from the Muscore test dataset and computed their corresponding rule labels to serve as targets. Subsequently, we generated 200 samples conditioned on each rule label. The rule labels for these generated samples were computed, and the loss between the generated rule label and the target was calculated. For pitch histogram and note density, we use MSE loss. For chord progression, we use 0-1 loss. The mean and standard deviation of this loss across the 200 samples are presented in Table \ref{tab:single_rule}.

We also compute OA between the generated samples and the data that comply with the desired rule. Specifically, we cluster 200 generated samples into 4 groups based on note density and compute OA within each group. The mean and standard deviations of OA across the 4 groups are also presented in Table \ref{tab:single_rule}.

\textbf{Composite Rule Guidance.}
We randomly selected 200 samples from the Muscore test dataset and compute their rule labels for each of the three rules under consideration. Then we generated 200 samples conditioned on all three rule labels simultaneously, with the intention that the generated samples adhere to all three rules concurrently. 

\textbf{Ablation Studies.}
For all the ablation studies, we follow the individual rule guidance set up and guide the diffusion model to generate music following given note density. The computational time is for generating 4 samples in a batch.
Regarding the quality metric, we randomly chose 200 samples from the test dataset and calculated the average OA similar to the process used for unconditional generation. This procedure was repeated five times to calculate the mean and standard deviation.

\textbf{Editing.}
To facilitate a comparison with MuseMorphase \cite{wu2023musemorphose}, we adopted their approach for computing the note density label. Initially, we randomly selected 200 samples from the Pop test dataset and calculated both vertical and horizontal note density vectors for each sample, using a window size of 1.28 seconds. Consequently, for each 10.24-second sample, we obtained 8 vertical and 8 horizontal note density values. We then flattened these note density vectors and categorized them into 8 bins, ensuring approximately an equal number of samples in each bin for both vertical and horizontal densities. During generation, we randomly chose a shift value of -1, 0, or 1 to adjust the note density classes of a sample, using the center value of the resultant bin as the target note density.

We also considered PolyDiff \cite{min2023polyffusion} as another baseline. In their approach, a new musical piece is generated based on the piano roll with basic chords extracted from the current piece, which is seen as an editing task since it creates a variation of the existing music with the same chord progression. In our framework, we extracted chords from the source music, added noise to the source, and then generated new music guided by the extracted chords.

For both baseline methods, we generated 8-bar music segments and extracted rule labels for each bar. In contrast, our method involved generating 10.24-second music segments and using a 1.28-second time window to extract rule labels, thereby aligning the number of rule labels across all methods.

In terms of similarity metrics, we calculated chroma and grooving similarity between the generated samples and their respective source samples, following the methodology outlined in \cite{wu2023musemorphose}.

\section{Training Surrogate Models for Music Rules}
\label{sec:appen_cls_train}
For classifier guidance \cite{dhariwal2021diffusion} and DPS-NN \cite{chung2022diffusion}, we need to train surrogate models to approximate various music rules. We used the DiT-S architecture in \cite{peebles2023scalable} as the backbone for classifiers \footnote{We use classifiers to refer to the surrogate models, even if they are not necessarily trained using a classification objective}. Following  the ViT approach \cite{dosovitskiy2020image}, we appended a class token to the latent codes and utilized a multi-layer perceptron (MLP) for the classification head. For rules like pitch histogram and note density, our classifier produces a vector of corresponding rules, and we train it using L2 loss. For chord progression, we incorporated two classifier heads: one to predict the key logits and the other for chord logits for each 128-length segment. We treated key and chord as categorical variables and trained the model using cross-entropy loss.

\begin{figure}[ht]
\centering
    	\includegraphics[width=\linewidth]{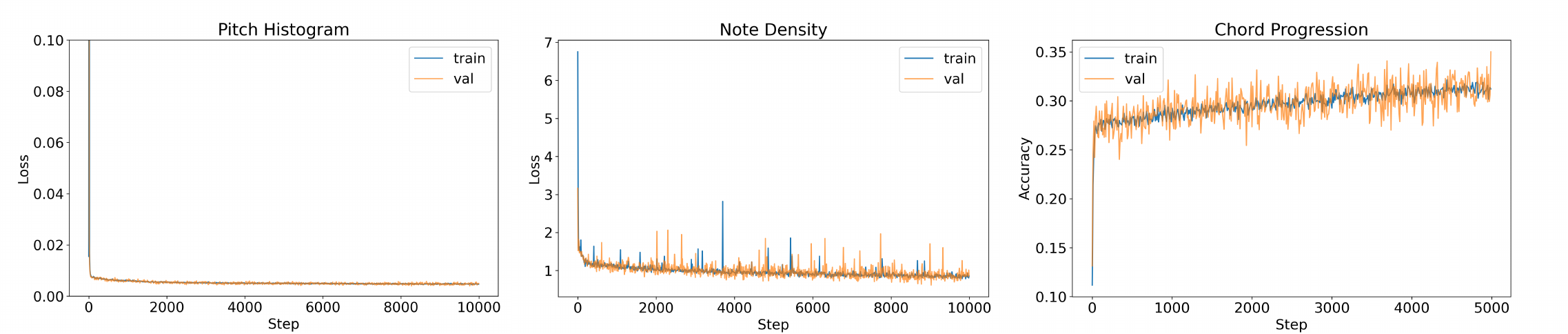}
        \caption{Training and validation curves of the classifiers trained on various rules.}
\label{fig:cls_curve}
\end{figure}

Figure \ref{fig:cls_curve} illustrates the training and validation loss/accuracy for the three rules under study. Notably, training a surrogate model for chord progression proved to be particularly challenging, with the final accuracy hovering around only $33\%$. This lower accuracy partly accounts for the diminished performance of rule-guided methods that depend on surrogate models. 

\section{Losses over Stochastic Control Guided Sampling Process}
\label{sec:appen_inter_loss}
\begin{figure}[ht]
\centering
    	\includegraphics[width=\linewidth]{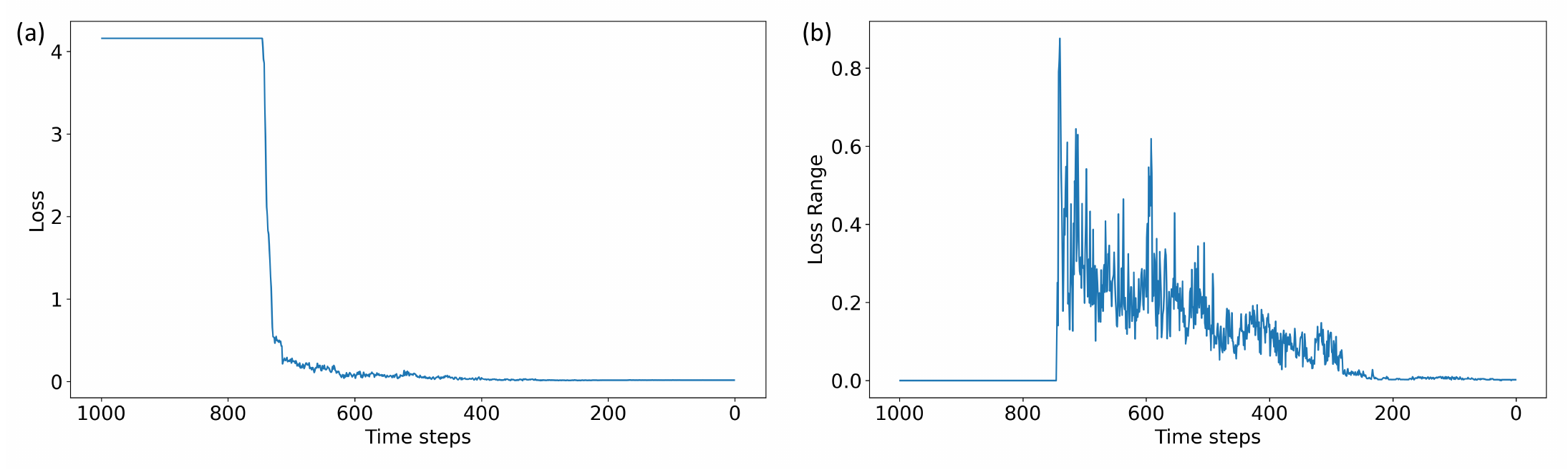}
        \caption{Best loss (a) and loss range (b) over stochastic control guided DDPM sampling on a representative sample with note density as the conditioning rule.}
\label{fig:inter_loss}
\end{figure}

We recorded the lowest loss and the variation in losses at each step throughout the sampling process of a representative sample using \ours, with note density as the conditioning rule. 
As depicted in Figure \ref{fig:inter_loss} (a), we observed that the loss remains consistent until approximately $t=750$. This early-stage constancy is attributed to the fact that, initially, the decoded piano rolls are essentially empty following the background thresholding, leading to a zero note density and, consequently, a stable loss. However, as the decoded piano rolls begin to populate, various realizations yield different note densities, resulting in a diversity of losses. By selecting the lowest loss at every step, we achieved a decrease in overall loss.

Figure \ref{fig:inter_loss} (b) illustrates the range of the losses at each step. The largest range occurs around $t=750$, the point where the piano roll starts to gain semantic meaning and the best loss drops drastically.
This suggests that applying guidance early, soon after the piano roll acquires semantic content, is crucial for successful guidance.

\section{More Ablation Studies}
\subsection{Unconditional Generation}
Our model is capable of generating samples reflective of the distributions from three distinct datasets. This is accomplished through classifier-free guidance \footnote{Despite this approach, we refer to it as `unconditional generation' because it does not involve rule-based guidance.} \cite{ho2022classifier}, with conditioning based on the specific dataset. We tuned the strength of the classifier-free guidance for each dataset and discovered optimal settings for achieving the highest music quality (Table \ref{uncond_gen_cfg_w}). Specifically, for the Maestro and Pop datasets, a guidance strength of $\omega=0$ yielded the best results. In contrast, for the Muscore dataset, setting $\omega=4$ proved to be most effective in enhancing musical quality.

\begin{table*}[h]
\centering
\resizebox{\textwidth}{!}{
\begin{tabular}{llcccccccc}
\toprule
Dataset &  $w$ &                    Used Pitch &                           IOI &                    Pitch Hist &                   Pitch Range &                      Velocity &                 Note Duration &                  Note Density &                           Avg \\
\midrule
\multirow{4}{*}{Maestro} &       0 &    $\mathbf{0.961 \pm 0.006}$ & $\underline{0.901 \pm 0.009}$ & $\underline{0.960 \pm 0.010}$ &    $\mathbf{0.963 \pm 0.006}$ &    $\mathbf{0.971 \pm 0.004}$ &    $\mathbf{0.910 \pm 0.012}$ &             $0.934 \pm 0.005$ &    $\mathbf{0.943 \pm 0.003}$ \\
 &       1 &             $0.948 \pm 0.002$ &    $\mathbf{0.915 \pm 0.011}$ &             $0.946 \pm 0.004$ &             $0.931 \pm 0.009$ &             $0.956 \pm 0.008$ &    $\mathbf{0.910 \pm 0.007}$ & $\underline{0.937 \pm 0.008}$ & $\underline{0.935 \pm 0.003}$ \\
 &       2 & $\underline{0.953 \pm 0.004}$ &             $0.878 \pm 0.006$ &    $\mathbf{0.964 \pm 0.003}$ & $\underline{0.946 \pm 0.006}$ & $\underline{0.963 \pm 0.008}$ &             $0.892 \pm 0.009$ &    $\mathbf{0.953 \pm 0.004}$ & $\underline{0.935 \pm 0.001}$ \\
 &       4 &             $0.925 \pm 0.005$ &             $0.891 \pm 0.008$ &             $0.940 \pm 0.007$ &             $0.934 \pm 0.011$ &             $0.958 \pm 0.010$ &             $0.890 \pm 0.010$ &             $0.932 \pm 0.005$ &             $0.924 \pm 0.001$ \\
\midrule
\multirow{4}{*}{Muscore} &       0 &             $0.941 \pm 0.003$ &             $0.890 \pm 0.019$ & $\underline{0.950 \pm 0.014}$ &             $0.946 \pm 0.006$ & $\underline{0.886 \pm 0.010}$ &    $\mathbf{0.922 \pm 0.011}$ &             $0.925 \pm 0.006$ & $\underline{0.923 \pm 0.003}$ \\
 &       1 & $\underline{0.956 \pm 0.008}$ &             $0.870 \pm 0.014$ &             $0.942 \pm 0.007$ & $\underline{0.962 \pm 0.009}$ &             $0.885 \pm 0.006$ & $\underline{0.911 \pm 0.007}$ & $\underline{0.934 \pm 0.007}$ & $\underline{0.923 \pm 0.003}$ \\
 &       2 &             $0.942 \pm 0.009$ & $\underline{0.899 \pm 0.014}$ &             $0.940 \pm 0.014$ &             $0.954 \pm 0.008$ &             $0.884 \pm 0.014$ & $\underline{0.911 \pm 0.007}$ &             $0.924 \pm 0.008$ &             $0.922 \pm 0.005$ \\
 &       4 &    $\mathbf{0.963 \pm 0.004}$ &    $\mathbf{0.915 \pm 0.009}$ &    $\mathbf{0.962 \pm 0.009}$ &    $\mathbf{0.964 \pm 0.004}$ &    $\mathbf{0.890 \pm 0.006}$ &             $0.900 \pm 0.012$ &    $\mathbf{0.946 \pm 0.005}$ &    $\mathbf{0.934 \pm 0.003}$ \\
\midrule
\multirow{4}{*}{Pop} &       0 &             $0.927 \pm 0.009$ &    $\mathbf{0.952 \pm 0.004}$ &             $0.969 \pm 0.002$ & $\underline{0.928 \pm 0.013}$ &    $\mathbf{0.948 \pm 0.003}$ &             $0.911 \pm 0.019$ &             $0.941 \pm 0.009$ &    $\mathbf{0.939 \pm 0.004}$ \\
     &       1 &             $0.921 \pm 0.004$ & $\underline{0.917 \pm 0.003}$ &    $\mathbf{0.975 \pm 0.003}$ &    $\mathbf{0.937 \pm 0.009}$ &             $0.939 \pm 0.007$ &    $\mathbf{0.926 \pm 0.015}$ &             $0.935 \pm 0.005$ & $\underline{0.936 \pm 0.003}$ \\
     &       2 & $\underline{0.929 \pm 0.008}$ &             $0.885 \pm 0.010$ & $\underline{0.970 \pm 0.005}$ &             $0.926 \pm 0.013$ &             $0.935 \pm 0.007$ &    $\mathbf{0.926 \pm 0.012}$ & $\underline{0.950 \pm 0.011}$ &             $0.932 \pm 0.003$ \\
     &       4 &    $\mathbf{0.933 \pm 0.011}$ &             $0.897 \pm 0.004$ &             $0.965 \pm 0.007$ &             $0.920 \pm 0.010$ & $\underline{0.940 \pm 0.007}$ &             $0.914 \pm 0.015$ &    $\mathbf{0.962 \pm 0.007}$ &             $0.933 \pm 0.005$ \\
\hline
\end{tabular}
}
\caption{Unconditional generation on three datasets with different classifier-free guidance strength.}
\label{uncond_gen_cfg_w}
\end{table*}

\subsection{Latent vs Pixel Space}
\label{sec:appen_pixel}
Our approach employed a latent diffusion model for symbolic music generation and compared its performance with a diffusion model trained in pixel space. The pixel space model was configured with a time resolution of 0.08 seconds per column in the piano roll, as opposed to the 0.01-second resolution in latent space. This choice was primarily driven by computational constraints; a 0.01-second resolution for a 10.24-second music piece would result in a piano roll of size $3 \times 128 \times 1024$, posing significant computational demands. In contrast, a 0.08-second resolution yields a more manageable size of $3 \times 128 \times 128$. For training the pixel space diffusion model, we utilized a standard U-Net backbone.

Table \ref{uncond_gen_pixel} presents a comparison of the models in the task of unconditional generation. An intriguing trend emerged: the latent space model excelled with complex, dynamic-rich music (e.g., Maestro), whereas the pixel space model showed superior performance with simpler music (e.g., Pop). The Muscore dataset, predominantly featuring classical sheet music, sits between Maestro and Pop in terms of complexity, and here, both models performed comparably. This observation aligns with the notion that time resolution has a less pronounced impact on the expressiveness of simpler music, making a lower resolution viable for training the diffusion model.

\begin{table*}[h]
\centering
\resizebox{\textwidth}{!}{
\begin{tabular}{llcccccccc}
\toprule
dataset &   method &                    Used Pitch &                           IOI &                    Pitch Hist &                   Pitch Range &                      Velocity &                 Note Duration &                  Note Density &                           Avg \\
\toprule
\multirow{2}{*}{Maestro} &  pixel &          $0.919 \pm 0.005$ &          $0.877 \pm 0.018$ & $\mathbf{0.983 \pm 0.005}$ &          $0.959 \pm 0.007$ &          $0.969 \pm 0.003$ &          $0.897 \pm 0.013$ &          $0.896 \pm 0.003$ &          $0.929 \pm 0.006$ \\
 & latent & $\mathbf{0.961 \pm 0.006}$ & $\mathbf{0.901 \pm 0.009}$ &          $0.960 \pm 0.010$ & $\mathbf{0.963 \pm 0.006}$ & $\mathbf{0.971 \pm 0.004}$ & $\mathbf{0.910 \pm 0.012}$ & $\mathbf{0.934 \pm 0.005}$ & $\mathbf{0.943 \pm 0.003}$ \\
\midrule
\multirow{2}{*}{Muscore} &  pixel &          $0.962 \pm 0.005$ &          $0.903 \pm 0.009$ & $\mathbf{0.965 \pm 0.009}$ & $\mathbf{0.964 \pm 0.007}$ & $\mathbf{0.893 \pm 0.007}$ & $\mathbf{0.928 \pm 0.011}$ &          $0.926 \pm 0.008$ & $\mathbf{0.934 \pm 0.005}$ \\
 & latent & $\mathbf{0.963 \pm 0.004}$ & $\mathbf{0.915 \pm 0.009}$ &          $0.962 \pm 0.009$ & $\mathbf{0.964 \pm 0.004}$ &          $0.890 \pm 0.006$ &          $0.900 \pm 0.012$ & $\mathbf{0.946 \pm 0.005}$ & $\mathbf{0.934 \pm 0.003}$ \\
 \midrule
\multirow{2}{*}{Pop} &  pixel & $\mathbf{0.935 \pm 0.011}$ & $\mathbf{0.957 \pm 0.004}$ & $\mathbf{0.976 \pm 0.003}$ & $\mathbf{0.952 \pm 0.004}$ &          $0.945 \pm 0.006$ & $\mathbf{0.935 \pm 0.011}$ & $\mathbf{0.946 \pm 0.013}$ & $\mathbf{0.949 \pm 0.005}$ \\
 & latent &          $0.927 \pm 0.009$ &          $0.952 \pm 0.004$ &          $0.969 \pm 0.002$ &          $0.928 \pm 0.013$ & $\mathbf{0.948 \pm 0.003}$ &          $0.911 \pm 0.019$ &          $0.941 \pm 0.009$ &          $0.939 \pm 0.004$ \\
\bottomrule
\end{tabular}
}
\caption{Comparing pixel vs latent space for unconditional generation.}
\label{uncond_gen_pixel}
\end{table*}

Table \ref{tab:single_rule_abla_pixel} shows the loss for individual rule guidance using the pixel space-trained diffusion model. Mirroring the findings in Table \ref{tab:single_rule}, \ours consistently achieved the lowest loss, underscoring its effectiveness in rule guidance. However, a noticeable decline in music quality (measured by OA) was observed for the model trained on pixel space, particularly in aspects like pitch histogram and note density (Table \ref{tab:single_rule_abla_pixel_quality}). This decline can be attributed to the nature of Gaussian noise addition in pixel space, which often results in random, musically nonsensical notes that nevertheless align with rule targets. Conversely, noise addition in latent space tends to induce more meaningful alterations, thereby preserving the higher music quality.

\begin{table}[h]
\centering
\resizebox{0.6\columnwidth}{!}{
\begin{tabular}{lccc}
\toprule
Method & Pitch Histogram $\downarrow$ & Note Density $\downarrow$ & Chord Progression $\downarrow$ \\
\midrule
No Guidance & $0.019 \pm 0.011$ & $2.367 \pm 2.933$ & $0.841 \pm 0.142$ \\
Classifier  & $0.020 \pm 0.015$ & $0.287 \pm 0.330$ & $0.783 \pm 0.208$ \\
DPS - NN    & $0.020 \pm 0.013$ & $0.615 \pm 1.188$ & $0.788 \pm 0.170$ \\
DPS - Rule  & $0.002 \pm 0.006$ & $2.349 \pm 3.425$ & - \\
\ours       & $\mathbf{0.0001 \pm 0.0008}$ & $\mathbf{0.103 \pm 0.570}$ & $\mathbf{0.344 \pm 0.212}$ \\
\bottomrule
\end{tabular}
}
\caption{Loss between the target and the generated attributes for individual rule guidance using the pixel-space trained diffusion model. }
\label{tab:single_rule_abla_pixel}
\end{table}

\begin{table}[h]
\centering
\resizebox{0.5\textwidth}{!}{%
\begin{tabular}{lccc}
\toprule
Model & Pitch Histogram $\uparrow$ & Note Density $\uparrow$ & Chord Progression $\uparrow$ \\
\midrule
Pixel &  $0.848 \pm 0.005$  &  $0.797 \pm 0.005$   & $\mathbf{0.892 \pm 0.009}$ \\
Latent &  $\mathbf{0.897 \pm 0.006}$  &  $\mathbf{0.880 \pm 0.003}$  &  $0.883 \pm 0.002$ \\
\bottomrule
\end{tabular}%
}
\caption{Comparison of Average Overlapping Area (OA) for individual rule guidance between diffusion models trained on pixel and latent space. OA is computed using the full dataset as reference.}
\label{tab:single_rule_abla_pixel_quality}
\end{table}

\subsection{Composite Rule Guidance}
\label{sec:appen_multiple_rule}
In the task of composite rule guidance, the allocation of suitable weights to each rule is crucial for effective rule-based guidance. Table \ref{tab:multi_rule_abla_weight} shows the performance associated with various weight assignments. Generally, we observed that amplifying the weight assigned to a specific rule tends to decrease the loss pertinent to that rule. However, excessively concentrating the weight on a single rule can lead to a deterioration in performance, as evidenced by the configuration with a 40-1-4 weight assignment with an overly heavy emphasis on chord progression (CP).

\begin{table}[h]
\centering
\resizebox{0.65\columnwidth}{!}{
\begin{tabular}{lcccc}
\toprule
Weight   & PH $\downarrow$    & ND   $\downarrow$   & CP  $\downarrow$    & OA full   $\uparrow$         \\
\midrule
40-1-1  &  $0.004 \pm 0.005$  & $0.218 \pm 0.243$   & $0.447 \pm 0.226$   &  $0.901 \pm 0.003$   \\
40-1-2  & $0.004 \pm 0.004$ & $\mathbf{0.215 \pm 0.193}$ & $\mathbf{0.392 \pm 0.206}$ &  $\mathbf{0.905 \pm 0.007}$   \\
40-1-4  & $0.004 \pm 0.004$  & $0.251 \pm 0.236$  & $0.418 \pm 0.216$ & $0.884 \pm 0.011$  \\
100-1-1 & $\mathbf{0.003 \pm 0.002}$ & $0.236 \pm 0.244$  & $0.434 \pm 0.229$ & $0.903 \pm 0.005$ \\
\bottomrule
\end{tabular}
}
\caption{Composite rule guidance using Classifier + \ours-4 with different weight on each rule. The weight column displays the weight in the order of PH, ND and CP.}
\label{tab:multi_rule_abla_weight}
\end{table}

Additionally, we investigated the impact of the sample count $n$ on composite rule guidance, as shown in Table \ref{tab:multi_rule_abla_bf}. The observed trend is consistent with that in individual rule guidance: using a greater number of samples at each step results in a lower loss. Another noteworthy observation is that combining \ours with other guidance methods (e.g., classifier guidance) and using a smaller sample count $n$ (such as 4) can yield better outcomes than using \ours alone with $n=16$. This improvement occurs because classifier guidance provides a coarse guidance signal, making it easier to identify advantageous directions based on these preliminary signals. As expected, the hybrid approach with a larger sample count $n=16$ achieves the lowest loss. Remarkably, the losses in this case are similar to those in individual rule guidance, despite being achieved simultaneously.

\begin{table}[h]
\centering
\resizebox{0.75\columnwidth}{!}{
\begin{tabular}{lccccc}
\toprule
Method   &  $n$    & PH $\downarrow$    & ND   $\downarrow$   & CP  $\downarrow$    & OA full   $\uparrow$         \\
\midrule
\ours &   16     & $0.014 \pm 0.009$     & $0.466 \pm 0.648$      & $0.446 \pm 0.205$      & $\mathbf{0.909 \pm 0.005}$      \\
DPS-NN + \ours & 4  &  $0.003 \pm 0.004$  & $0.392 \pm 0.612$   &  $0.486 \pm 0.270$  &  $0.826 \pm 0.005$   \\
Classifier + \ours & 4  &  $0.004 \pm 0.005$  & $0.218 \pm 0.243$   & $0.447 \pm 0.226$   &  $0.901 \pm 0.003$   \\
DPS-NN + \ours & 16 &  $\mathbf{0.002 \pm 0.007}$  &  $0.238 \pm 0.531$  &  $0.313 \pm 0.231$  &  $0.844 \pm 0.007$   \\
Classifier + \ours & 16  & $0.003 \pm 0.005$ & $\mathbf{0.148 \pm 0.203}$ & $\mathbf{0.284 \pm 0.197}$ &  $0.894 \pm 0.007$   \\
\bottomrule
\end{tabular}
}
\caption{Effect of number of samples $n$ on composite rule guidance.}
\label{tab:multi_rule_abla_bf}
\end{table}

\section{Rule-Guided Generation Survey}
\label{sec:rule-survey}
\subsection{Details}
To evaluate the rule alignment of our approach SCG compared to two baseline methods, we designed a listening test. We extracted three musical rules (Pitch Histogram, Note Density, and Chord Progression) from various segments in our dataset. Next, we created music samples lasting 10.24 seconds each, adhering to these three rules. We produced four samples for each guiding method, including our own, resulting in a total of 12 samples.

\begin{figure}[h!]
     \centering
     \begin{minipage}[t]{0.3\linewidth}
         \centering
         \includegraphics[width=0.9\linewidth]{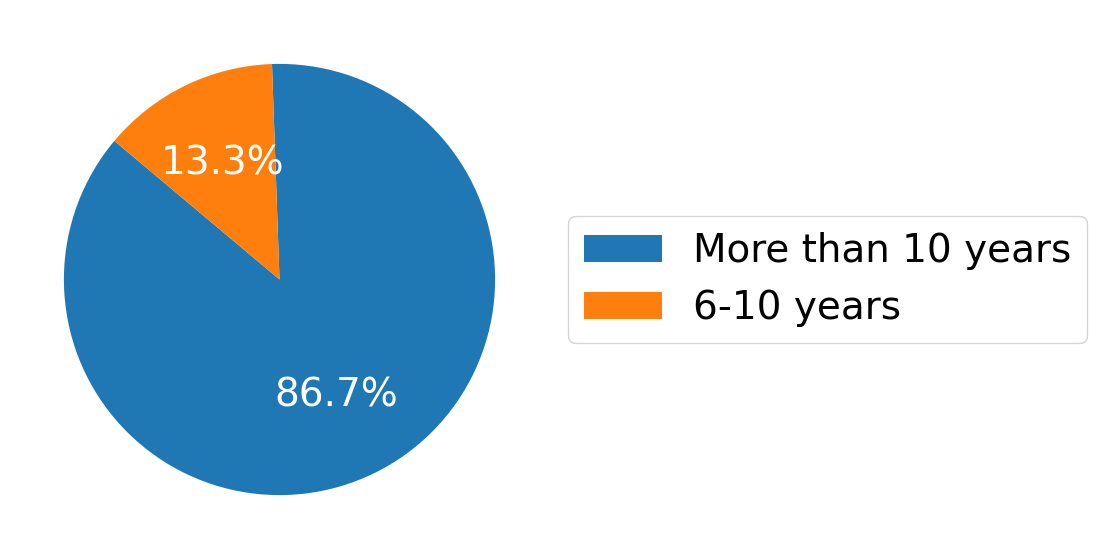}
         \caption{Years of playing music}
         \label{fig:survey-q1}
     \end{minipage}
     \hfill
     \begin{minipage}[t]{0.3\linewidth}
         \centering
         \includegraphics[width=0.9\linewidth]{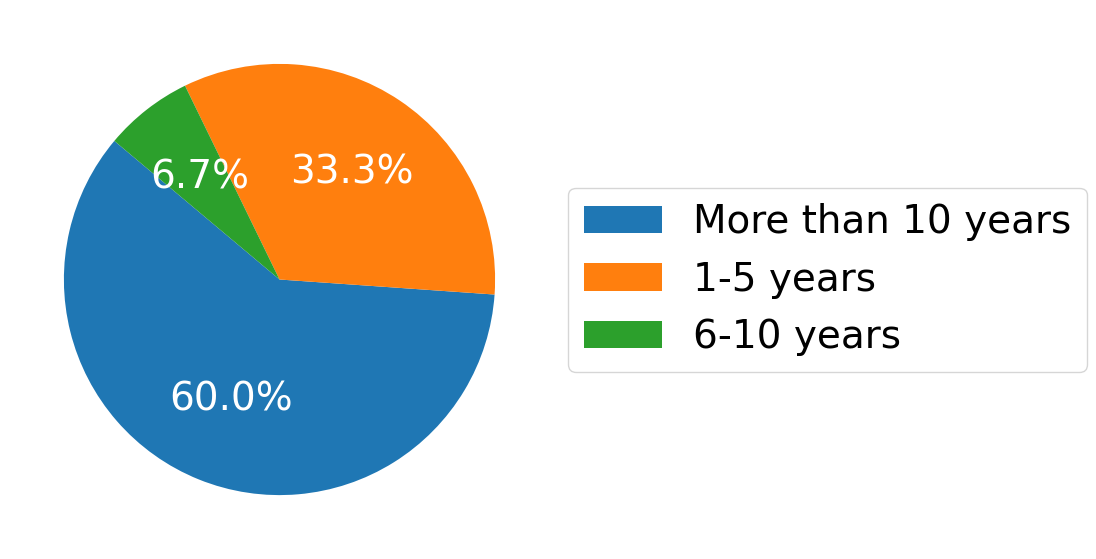}
         \caption{Years of formal music study}
         \label{fig:survey-q2}
     \end{minipage}
     \hfill
     \begin{minipage}[t]{0.3\linewidth}
         \centering
         \includegraphics[width=0.77\linewidth]{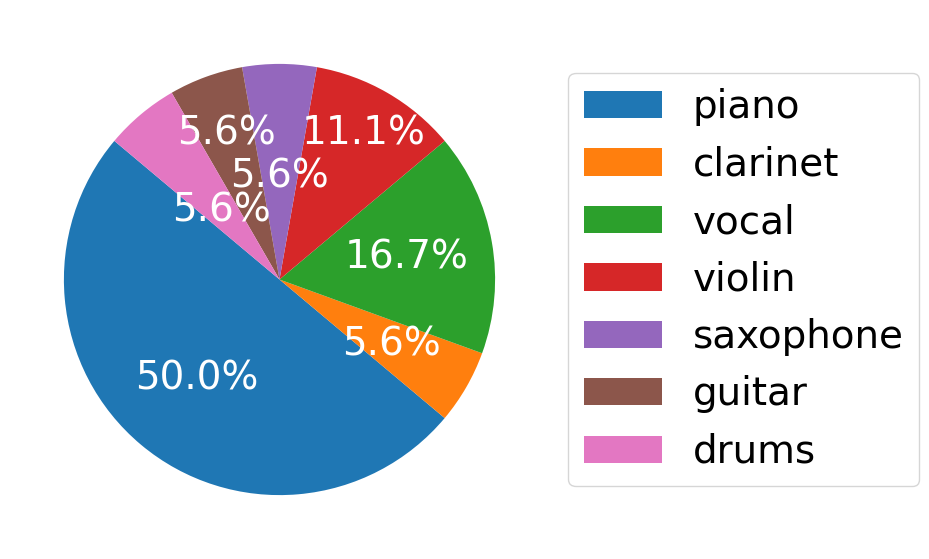}
         \caption{Instruments of participants}
         \label{fig:survey-q3}
     \end{minipage}
     \hfill
\end{figure}

We recruited 15 participants with substantial musical experience for our survey. We gathered information on their musical backgrounds, including the number of years they have been playing music, their years of formal music education, and the instruments they play. A significant portion of the participants have over 10 years of experience in both playing music and formal musical study, as shown in figures \ref{fig:survey-q1} and \ref{fig:survey-q2}. Figure \ref{fig:survey-q3} displays a diverse range of instrument expertise among participants, with a notable prevalence of piano players, aligning with our model's focus on piano music. This diversity and level of experience make the participants well-suited for analyzing the music's rule alignment and musicality.

The dimension we evaluate about the sample quality are rule alignment, creativity, coherence and overall rating. Question 1-3 are about rule alignment, which evaluates the performance of guidance. Creativity refers to whether samples are musically interesting or not. For example, if one segment is static, then such sample is not creative. Coherence refers to whether the samples align with basic musical common knowledge. For instance, if one segment contains many random notes, then such sample sounds chaotic and it is not coherent to human's sense of good music. The overall rating is evaluated by participants, where they give a score solely based on their preference to the samples. 

Each evaluation dimension is rated on a scale up to 5 points. For rule alignment, a Likert scale is used, where "completely unaligned" is rated as 1 and "perfectly aligned" as 5. The average score from questions 1-3 determines the rule alignment score. Creativity is initially scored out of 3 points, which is then normalized to a 5-point scale. The average of questions 4 and 5 calculates the creativity score. In question 4, "No" scores 1 point, "Maybe" 2 points, and "Yes" 3 points. For question 5, both "Too Simple" and "Too Complex" score 1 point, while "Moderate" scores 3 points.

For coherence, the maximum score is 4 points, later normalized to 5 points. The average from questions 6-8 gives the coherence score. In question 6, "Many" errors score 1 point, "Some" 2 points, "A Few" 3 points, and "None" 4 points. In question 7, "Mainly incoherent harmonic motion" scores 1 point, "Mainly incoherent harmonic motion with some coherence" 2 points, "Mainly coherent harmonic motion with some incoherence" 3 points, and "Coherent harmonic motion" 4 points. Question 8 uses a tailored scoring system to fit the 4-point scale: "Poor" is valued at 4/3 points, "Moderate" at 8/3 points, and "Highly Engaging" at 4 points.

The overall rating also utilizes a Likert scale, with "Poor" equating to 1 point, "Fair" 2 points, "Good" 3 points, "Very Good" 4 points, and "Excellent" 5 points.

Regarding the music rules (Pitch Histogram, Note Density, and Chord Progression), we generated the entire sample conditioned on pitch histogram. To assess note density and chord progression, the music segment is divided into 8 equal-length segments or windows. We then analyze and compute the note density and chord progression within each of these windows. The survey consists 9 questions in total, investigating SCG's rule alignment and sample's musicality. 

\subsection{Survey}
The first three questions of the survey are studying the rule alignment of guidance mechanisms. The answer of these three questions are all classified into 5 categories instead of binary choices because rule alignment can be effective for parts of the music sample. For example, given a 10 second music sample, the first 5 seconds of the sample has the perfect alignment, and the last 5 seconds of the sample does not align with provided rules at all. In this case, only binary classification on how effective the guidance is would not be enough to distinguish such sample. Thus, we construct 5 options instead of binary choices.

\textbf{Question 1: } On a scale of 1 to 5, how well does the pitch histogram in the sample music match the provided histogram? (1 indicating the least alignment, with 5 indicating the most alignment)

The options are: \begin{itemize}
    \item Completely unaligned (1): The pitch histogram in the sample music is completely different from the provided pitch histogram.
    \item Somewhat unaligned (2): The pitch histogram in the sample music is somewhat different from the provided pitch histogram, with a small portion of the segment aligned.
    \item Moderately aligned (3): The pitch histogram in the sample music is somewhat aligned with the provided pitch histogram, with a small portion of the segment not aligned.
    \item Mostly aligned (4): The pitch histogram in the sample music is mostly aligned with the provided pitch histogram.
    \item Perfectly aligned (5): The pitch histogram in the sample music is perfectly aligned with the pitch histogram.
\end{itemize}

Besides the generated, we show the participants the image of pitch histogram that are used for guidance. Note that Pitch histogram is the distribution of notes. Question 1 focuses on the alignment of pitch histogram, which means whether distribution of notes in the given sample follows the given pitch histogram. The question directly evaluates how effective the conditioning on pitch histogram is.

\textbf{Question 2: } On a scale of 1 to 5, how would you rate the alignment of the note density of the sample music compared to the note density provided in the above youtube video? (1 being the least aligned, 5 being the most aligned, take a look at the piano roll image would be a good idea)

The options are: \begin{itemize}
    \item Completely unaligned (1): The note density in the sample music significantly differs from the provided music segment, leading to a large disparity in musical texture and pacing.
    \item Somewhat unaligned (2): The note density in the sample music somewhat differs from the provided music segment, causing a noticeable disparity in musical texture and pacing.
    \item Moderately aligned (3): The note density in the sample music is somewhat aligned with the provided note density, but with a perceptible mismatch in musical flow and rhythm.
    \item Mostly aligned (4): The note density of the sample music closely matches the provided note density, with slight differences in how notes are spaced and arranged.
    \item Perfectly aligned (5): The note density of the sample music aligns perfectly with the provided note density, reflecting a very similar density pattern in the distribution of notes.
\end{itemize}

Question 2 evaluates the alignment of note density. Note density refers to the frequency and distribution of musical notes in a piece, indicating how many notes occur over a specific time or within a certain section of the music. In other words, note density reflects the texture and pacing in music segments. Thus, under a successful guidance of such rule, the texture and pacing of generated samples would be similar to the density pattern of corresponding segments in the distribution of notes. 

\textbf{Question 3: } On a scale from 1 to 5, how well does the chord progression in the sample music match the provided chord progression, focusing on their functional harmony and general sequence rather than specific chord inversions. (1 indicates minimal alignment, with pronounced differences in chord progression, 5 signifies complete alignment, with the chord progressions being very similar or identical)

The options are: \begin{itemize}
    \item (Completely Unaligned): The bass line of the chord progression in the sample music significantly deviates from the provided progression, resulting in a clear disparity in harmonic structure and musical direction.
    \item (Somewhat Unaligned): Observable differences in the chord progression between the sample music and the provided example lead to a discordant sound and a disrupted musical flow.
    \item(Moderately Aligned): The chord progression in the sample music is somewhat consistent with the provided progression, with only minor discrepancies in the sequence or harmony.
    \item (Mostly Aligned): The chord progression in the sample closely mirrors the provided progression, with only negligible variations that don’t substantially affect the overall harmonic continuity.
    \item (Perfectly Aligned): The chord progression in the sample music perfectly matches the provided progression, ensuring a cohesive and harmonious harmonic structure throughout.
\end{itemize}

Chord progression guides the music segment sounding more reasonable. Such questions asks about the alignment of chords, where effectively evaluates the controllable generation based on given chord progression. 

The subsequent five questions explore the musicality of the generated samples, encompassing both creativity and coherence. The primary aim of rule-based guidance is to enhance the auditory appeal of the samples, making them more enjoyable to listeners. A generation is not considered successful if it fails to be aesthetically pleasing, regardless of achieving perfect alignment with all three specified rules.

Questions 4 and 5 focus on evaluating the creativity of the music sample.

\textbf{Question 4: } Do you like the music based on your personal taste?

The options are: \begin{itemize}
    \item Yes
    \item No
    \item Maybe
\end{itemize}

Questions 4 directly asks whether the participants like the music based on their personal taste. The evaluation from listeners with substantial musical experience illustrates the quality of model generation.

\textbf{Question 5: } What do you think of the complexity of this music?

The options are: \begin{itemize}
    \item Too Simple
    \item Moderate
    \item Too Complex
\end{itemize}

Question 5 assesses the complexity of the music, indicating that a moderate level of complexity is optimal. Music that is either too simple or too complex is considered to detract from the quality of the sample.

Questions 6 to 9 are designed to assess the coherence of the music sample.

\textbf{Question 6: } How many elements in the sample that seem out of place or random?

The options are: \begin{itemize}
    \item None
    \item A Few
    \item Some
    \item Many
\end{itemize}

Question 6 aims to evaluate on the generation quality of the model. Because the sample is composed by the model instead of human, such sample might have random notes. The random elements would break the entity of the music segment and the pleasure of listening for participants.

\textbf{Question 7: } How coherent do you find the harmony in the excerpt to be?

The options are: \begin{itemize}
    \item Coherent harmonic motion
    \item Mainly coherent harmonic motion with some incoherence
    \item Mainly incoherent harmonic motion with some coherence
    \item Mainly incoherent harmonic motion
\end{itemize}

Question 7 assesses the harmonic coherence of the generated sample, specifically evaluating if the music segment appears harmonically random or structured. 

\textbf{Question 8: } How would you rate the appropriateness and engagement of the texture in the sample music, considering the layers and how they combine? 

The options are: \begin{itemize}
    \item Highly Engaging
    \item Moderate
    \item Poor
\end{itemize}

Question 8 examines the texture of the sample music, focusing on whether the music presents an overly complex or disordered structure.

Question 9 solicits the overall rating of the music, where participants rate the sample according to their personal preferences.

\textbf{Question 9: } Overall Rating: On a scale of 1 to 5, how would you rate this sample? (1 being lowest, 5 highest)

The options are: \begin{itemize}
    \item Poor: The music lacks appeal in many aspects and does not engage the listener.
    \item Fair: The music has some redeeming qualities but falls short in several areas.
    \item Good: The music is enjoyable and well-composed, though it may have a few minor flaws.
    \item Very Good: The music is engaging and impressive, showing high levels of creativity and skill.
    \item Excellent: The music is outstanding in all respects, offering a deeply satisfying and memorable listening experience.
\end{itemize}

\end{document}